\documentclass[11pt,notitlepage,amsmath,amssymb,aps,prl,nofootinbib]{revtex4-2}  

\usepackage{times}
\usepackage{graphicx}  
\usepackage{float}
\usepackage{dcolumn}   
\usepackage{bm}        
\usepackage{slashed}   
\usepackage{simplewick} 
\usepackage{verbatim}  
\usepackage{color} 
\usepackage{appendix} 
\usepackage{subfigure}

\graphicspath{{figure/}}

\newcounter{JW}

\usepackage[bookmarks=true,colorlinks=true,citecolor=red,linkcolor=blue,unicode=true]{hyperref}

\newcounter{YJC}

\begin{document}


\title{QCD on Rotating Lattice with Staggered Fermions}
\author{Ji-Chong Yang}
\email{yangjichong@lnnu.edu.cn}
\affiliation{Department of Physics, Liaoning Normal University, Dalian 116029, China,}
\affiliation{Center for Theoretical and Experimental High Energy Physics, Liaoning Normal University, Dalian 116029, China.}
\author{Xu-Guang Huang}
\email{huangxuguang@fudan.edu.cn}
\affiliation{Physics Department and Center for Particle Physics and Field Theory, Fudan University, Shanghai 200438, China,}
\affiliation{Key Laboratory of Nuclear Physics and Ion-beam Application (MOE), Fudan University, Shanghai 200433, China,}
\affiliation{Shanghai Research Center for Theoretical Nuclear Physics, Natural Science Foundation of China and Fudan University, Shanghai 200438, China}

\begin{abstract}
We investigate the finite-temperature quantum chromodynamics (QCD) on a rotating lattice with $N_f=2+1$ staggered fermions and the projective plane boundary condition. We observe a negative rotational rigidity (defined in the main text) and a negative quark spin susceptibility associated with the chiral vortical effect. In contrast to most of the effective model predictions, we find that the chiral condensate decreases and the Polyakov loop increases with imaginary rotation, implying a rotational catalysis of chiral symmetry breaking and confinement by real rotation. We determine the phase boundaries for both chiral and confinement-deconfinement phase transitions on the $\Omega_I$-$T$ plane, where $\Omega_I$ is the imaginary angular velocity.
\end{abstract}

\date{\today}

\maketitle

{\bf Introduction. ---} Fast rotation plays an important role in many quantum chromodynamics (QCD) systems. For instance, some neutron stars may rotate at angular velocities close to their Keplerian frequencies, which can affect their evolution, structure, magnetic fields, and stability~\cite{Paschalidis:2016vmz}. In relativistic heavy-ion collisions at the RHIC and LHC, very strong fluid vorticity (i.e., local angular velocity of the fluid cell) can be generated by the large angular momentum of the colliding nuclei~\cite{Deng:2016gyh,Jiang:2016woz}. The rotation or fluid vorticity can significantly influence the quark-gluon plasma and induce novel spin-related quantum phenomena, such as the chiral vortical effect (CVE)~\cite{Erdmenger:2008rm,Banerjee:2008th,Landsteiner:2011cp,Huang:2018aly}, which is the generation of vector or axial currents along fluid vortex, and the spin polarization of hyperons and spin alignment of vector mesons which have been observed recently~\cite{STAR:2017ckg,STAR:2022fan}.

Recently, the effect of rotation on the QCD phase structure has been extensively investigated~\cite{Chen:2021aiq}. Effective models have shown that rotation, at finite temperature, density, and magnetic field, acts like an effective chemical potential~\cite{Chen:2015hfc,Jiang:2016wvv,Chernodub:2016kxh}. (It has also been suggested that uniform rotation is thermodynamically invisible at zero temperature, density, and magnetic field~\cite{Vilenkin:1980zv,Ebihara:2016fwa,Chen:2017xrj}.) Consequently, rotation tends to reduce the chiral condensate and the chiral critical temperature $T_\chi$~\cite{Chen:2015hfc,Jiang:2016wvv,Chernodub:2016kxh,Chernodub:2017ref,Wang:2018sur,Wang:2018zrn,Wang:2019nhd,Zhang:2020jux,Sadooghi:2021upd,Chen:2022mhf,Chen:2023cjt}. Novel rotation-induced pion condensate may also emerge~\cite{Huang:2017pqe,Liu:2017spl,Zhang:2018ome,Chen:2019tcp,Cao:2019ctl,Nishimura:2020odq}. The effect of rotation on deconfinement phase transition remains controversial. Holographic QCD models~\cite{Chen:2020ath,Braga:2022yfe,Yadav:2022qcl,Zhao:2022uxc}, hadron resonance gas model~\cite{Fujimoto:2021xix}, and perturbative calculation~\cite{Chen:2022smf} suggest a decrease of deconfinement temperature $T_d$ by rotation. Other numerical simulations for pure gluons indicate mixed confinment-deconfinment phases~\cite{Chernodub:2022veq}. However, lattice simulations of $SU(3)$ gluondynamics appear to support an enhancement of confinement by rotation~\cite{Braguta:2020biu,Braguta:2021jgn}.

This paper aims to provide a comprehensive study of hot QCD under rotation using lattice simulations with $N_f=2+1$ staggered fermions. The lattice QCD approach to the rotating system in the quenched approximation was first introduced in Ref.~\cite{Yamamoto:2013zwa} and subsequent studies of the $SU(3)$ pure Yang-Mills case were conducted in Refs.~\cite{Braguta:2020biu,Braguta:2021jgn}. The rotation is implemented by simulating a rest state in a rotating frame, which is equivalent to simulating a rotating state in an inertial frame (See Appendix). In this work, we construct the lattice action following Ref.~\cite{Yamamoto:2013zwa} but with dynamic staggered quarks and a projective plane boundary condition.

{\bf \label{sec:model}Formulation. ---} The QCD Lagrangian in a frame rotating about the $z$-axis at a constant real angular velocity $\Omega$ is $\mathcal{L}_{\rm QCD}=\mathcal{L}_{\rm G} + \mathcal{L}_{\rm F}$, where
\begin{equation}
\begin{split}
&\mathcal{L}_{\rm G}=-\frac{1}{2g^2_{s}}g^{\mu\nu}g^{\rho\sigma }{\rm tr} \left(F_{\mu\rho}F_{\nu\sigma}\right),\\
&\mathcal{L}_{\rm F}=\bar{q}\left[i\gamma _{\mu}\left(\partial _{\mu}+iA_{\mu}+\Gamma _{\mu}\right)-m\right]q ,
\end{split}
\label{eq.2.1}
\end{equation}
and $A_\mu$ and $q$ are gluon and quark fields, and $m={\rm diag}(m_l, m_l, m_s)$ is the mass matrix for $N_f=2+1$ flavors. The Lorentzian-signature metric and the spin connection $\Gamma_\mu$ are given by
\begin{equation}
\begin{split}
&(g_{\mu\nu})=\left(\begin{array}{cccc} 1-r^2\Omega^2 & y\Omega & -x\Omega & 0 \\ y\Omega & -1 & 0 & 0 \\ -x\Omega & 0 & -1 & 0 \\ 0 & 0 & 0 & -1 \end{array}\right),\\
&\Gamma _{\mu}=-\frac{i}{4}\sigma ^{ab}w_{\mu ab},\;\;w_{\mu ab}=g_{\alpha \beta}e_a^{\alpha}(\partial _{\mu}e_b^{\beta}+\Gamma ^{\beta}_{\mu\nu}e_b^{\nu}),
\end{split}
\label{eq.2.2}
\end{equation}
where $r=\sqrt{x^2 + y^2}$ is the transverse distance, $\Gamma ^{\mu}_{\alpha\beta}$ is the Christoffel symbol, $\sigma^{ab}\equiv i [\gamma^a, \gamma^b]/2$, and $e_a^{\mu}$ is the vierbein with $e_0=(1,y\Omega,-x\Omega,0)$, $e_1=(0,1,0,0)$, $e_2=(0,0,1,0)$, and $e_3=(0,0,0,1)$. The Euclidean action is obtained by the Wick rotation which, however, generates a complex Euclidean-signature metric and causes a ``sign problem'' in Monte Carlo simulations. To avoid this sign problem, we follow Ref.~\cite{Yamamoto:2013zwa} and replace $\Omega$ by an imaginary rotation $\Omega_I$, $\Omega \rightarrow i \Omega _I$, in our simulation.

We use the same discretization of the gauge action $S_{\rm G}$ as in Ref.~\cite{Yamamoto:2013zwa} (see also Appendix). We discretize the quark action $S_{\rm F}$ using staggered quarks:
\begin{equation}
\begin{split}
&S_{\rm F}=\sum _n \Big\{\sum _{\mu}\sum _{\delta = \pm \mu}\bar{\psi} (n)V(n,n+\hat{\delta})\psi (n+\hat{\delta})\\
&\qquad+2a m \bar{\psi} (n)\psi (n)+\frac{\Omega_I}{4} \left(b_{x,y}-b_{y,x}\right)\\
&\qquad+\frac{a\Omega_I}{8}\sum _{s_{x,y,\tau}=\pm 1}s_{\tau}\eta _{s_{\tau}\tau}(n)\eta _{xy}(n)\\
&\qquad\times \bar{\psi} (n)U\Big(n,n+\sum _{i=x,y,\tau}s_i \hat{i}\Big)\psi \Big(n+\sum _{i=x,y,\tau}s_i \hat{i}\Big)\Big\},
\end{split}
\label{eq.2.4}
\end{equation}
where $\psi$ is the staggered quark field~\cite{Rothe:1992nt}, $a$ is the lattice spacing, $\beta \equiv 2N_c / g_{s}^2$ is the lattice coupling ($N_c=3$), $\hat{\mu}$ is the dimensionless unit vector along $\mu$-axis, and $\eta _{\mu}(n)$ is defined on links by $\eta _{\mu}(n)=(-1)^{\sum _{\nu<\mu}n_{\nu}}$ and $\eta _{-\mu}(n)=-\eta _{\mu}(n-\hat{\mu})$. In the above, $U(n_1,n_2)$ is the average of the shortest Wilson lines connecting $n_1$ and $n_2$, $V(n_1,n_2)\equiv U(n_1,n_2)\eta(n_1,n_2)$ with $\eta (n_1,n_2)$ the product of $\eta _{\mu}$ along the shortest path connecting $n_1$ and $n_2$, and
\begin{equation}
\begin{split}
&b_{i,j}\equiv \sum _{s_{\tau,i}=\pm 1} s_{\tau}s_i n_j \bar{\psi}(n-s_i \hat{i})V\big(n-s_i \hat{i},n+s_{\tau}\hat{\tau}+s_i \hat{i}\big)\\
&\qquad\times \psi (n+s_{\tau}\hat{\tau} + s_i \hat{i}),\\
&\eta _{xy}(n)=\eta _x(n)\eta _y(n).
\end{split}
\label{eq.2.5}
\end{equation}

\begin{figure}[!htbp]
\begin{center}
\includegraphics[width=0.18\textwidth]{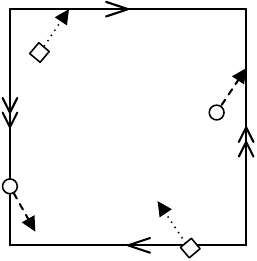}
\caption{\label{fig:p2a}Illustration of a two-dimensional projective plane. The edges with matching arrows are identified along the arrow directions. 
}
\end{center}
\end{figure}
A uniformly rotating system must be finite to preserve causality. Therefore, the boundary condition along the directions perpendicular to the rotation axis is crucial. We use a special periodic boundary condition that makes the ${x y}$-plane of the lattice a projective plane, as shown in Fig.~\ref{fig:p2a}. This boundary condition has two advantages. First, it ensures a smooth gauge action at the boundaries and reduces the rotation-independent effects from the boundaries~\cite{Tiburzi:2013vza}. Second, unlike the Dirichlet boundary condition, it allows rotational spinor eigenstates to exist~(see Appendix).


We implement the Monte Carlo simulation on $N_x^3\times N_\tau=12^3\times 4$ and $12^3\times 6$ lattices with $N_f=2+1$ dynamic staggered quarks and various $\beta$. The bare masses are $m_l\approx 20\;{\rm MeV}$ for $u, d$ quarks and $m_s=5m_l$ for the strange quark. We set the chemical potential to zero. Other lattice parameters are listed in the Appendix.

{\bf Rotational rigidity and spin susceptibility. ---}
We first consider the response of the QCD matter to rotation, namely, the generation of angular momentum and fermionic current (i.e., the CVE) by rotation. The angular momentum in QCD draws a lot of attention because of the proton spin puzzle~\cite{Ji:2020ena} and the observation of global spin polarization/alignment of hadrons in heavy-ion collisions~\cite{STAR:2017ckg,STAR:2022fan}. For the chemical potential is zero, only the axial CVE is present which is closely related to the quark spin contribution to the angular momentum. We concentrate on the case of $N_{\tau}=6$.

\begin{figure}[!htbp]
\begin{center}
\includegraphics[width=0.35\textwidth]{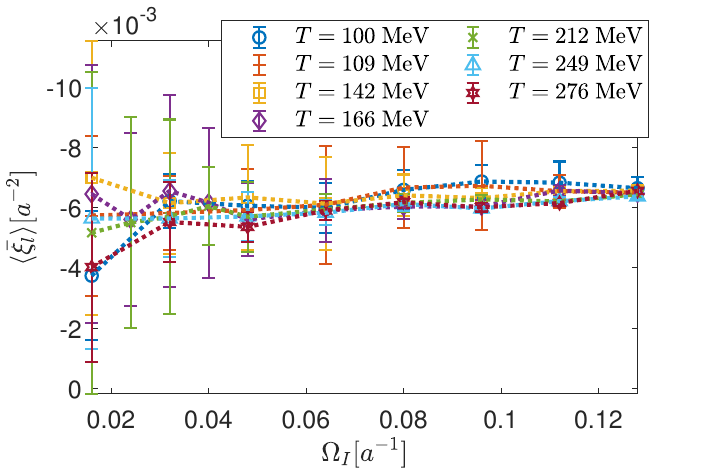}
\includegraphics[width=0.35\textwidth]{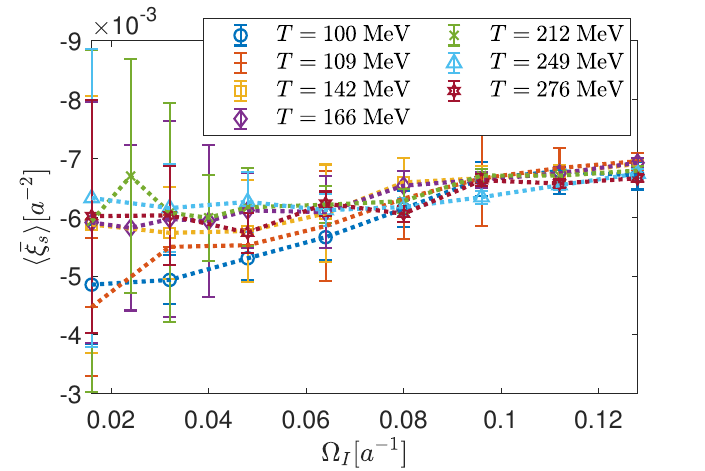}
\caption{\label{fig:xi} The space-averaged spin susceptibilities $\bar\xi _{l,s}$ as functions of $\Omega_I$.}
\end{center}
\end{figure}

The angular momentum (density) $\bf{J}$ of QCD can be decomposed into different components in different ways. We use  Ji's decomposition~\cite{Ji:1996ek},  ${\bf J}={\bf J}_G+\sum_f \left ({\bf s}_f+{\bf L}_f\right)$, in which
\begin{equation}
\begin{split}
{\bf J}_G&= \sum _a \;{\bf x} \times \left( {\bf E}^a \times {\bf B}^a\right),\\
{\bf s}_f&=  q_f ^{\dagger}\frac{\bf \Sigma}{2}  q_f, \\
{\bf L}_f&= \frac{1}{i}\; q_f^{\dagger}{\bf x} \times ({\bm \partial}-i{\bf A}) q_f,
\end{split}
\label{eq.4.3}
\end{equation}
where ${\bf\Sigma}={\rm diag}({\bm\sigma}, {\bm\sigma})$ with ${\bm\sigma}$ the Pauli matrices. In this decomposition, ${\bf J}_G$ is the gluon angular momentum,  ${\bf s}_f$ is the quark spin of flavor $f$, and ${\bf L}_f$ is the quark orbital angular momentum; they are all gauge invariant. When $r$ is small, it is known that the radial distributions of $J^z_G(r)$ and $L^z_f(r)$ are approximately quadratic, while $s^z_f(r)$ is insensitive to $r$~\cite{Yamamoto:2013zwa}. Under imaginary rotation, the angular momentum is also imaginary. We thus compute the ratios $\xi_f\equiv s^z_f(r)/\Omega$, $\rho_{J_G} \equiv J^z_G(r)/(\Omega\, r^2)$, $\rho_{L_f} \equiv L^z_f(r)/(\Omega\, r^2)$ on lattice which are real before and after Wick rotation. They characterize the strengths of different components of ${\bf J}$ in response to uniform rotation. We call $\rho$'s the {\it rotational rigidities} and $\xi_f$ the {\it spin susceptibility} of flavor $f$~\footnote{We use $\xi_f$ instead of the more natural definition $\chi_{\Omega}^f\equiv \partial s_f^z/\partial\Omega$ for spin susceptibility because $\xi_f$ can be more accurately measured on lattice. They almost coincide as $s_f^z$ is roughly linear in $\Omega$ within the computational error (see Fig.~\ref{fig:xi}).}. Note that, for a non-relativistic system, such defined $\rho$ reduces to the mass density.

\begin{figure}[!htbp]
\begin{center}
\includegraphics[width=0.35\textwidth]{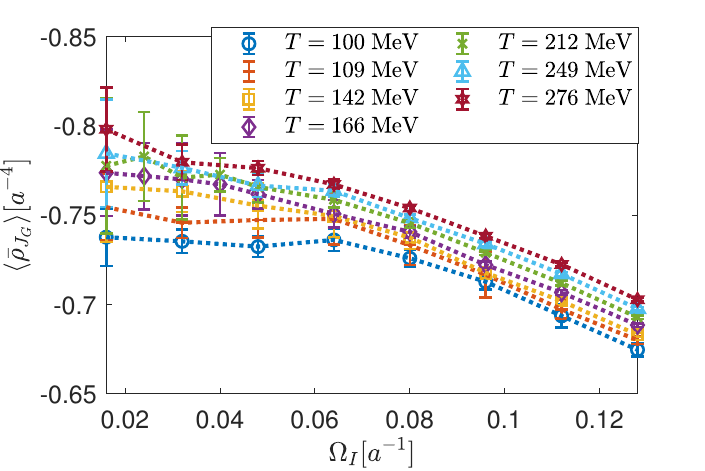}
\includegraphics[width=0.35\textwidth]{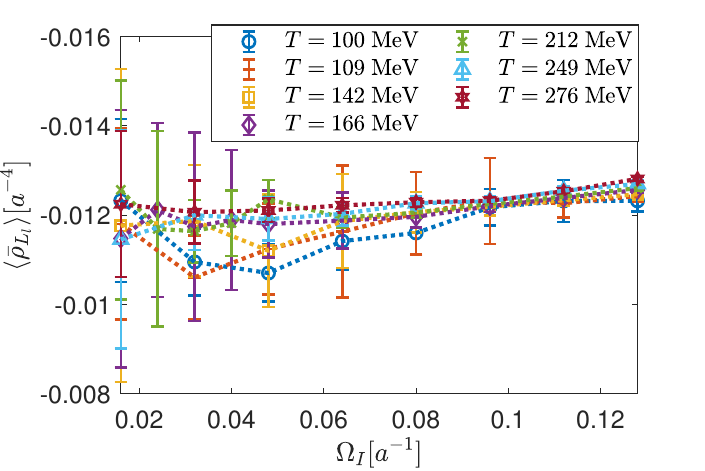}
\caption{\label{fig:rho} The space-averaged rotational rigidities $\bar\rho _{J_G}$ and $\bar\rho _{L_{l}}$ as functions of $\Omega_I$.}
\end{center}
\end{figure}

We average $\langle \rho _{J_G}\rangle$, $\langle \rho _{L_{f}}\rangle$ and $\langle \xi _{l,s}\rangle$ over the whole lattice and show the results in Figs.~\ref{fig:rho} and \ref{fig:xi}. Within the statistical error, $\langle\bar\rho_{L_s}\rangle$ (space-averaged quantities are denoted with an overbar) is the same as $\langle\bar\rho_{L_l}\rangle$, and therefore only $\langle\bar\rho_{L_l}\rangle$ is shown. We find that, over a large temperature regime $T\approx100$-$300$ MeV, $\langle\bar\rho_{J_G}\rangle$ is negative with magnitude slightly decreasing with the growth of $\Omega_I$. On the other hand, $\langle\bar\rho _{L_{l,s}}\rangle$ are also negative but insensitive to $\Omega_I$. The negativity may indicate a {\it thermodynamic instability} of QCD against uniform rotation. A recent lattice simulation for $SU(3)$ pure gluons obtains a negative moment of inertial which is equivalent to the negativity of $\langle\bar\rho_{J_G}\rangle$~\cite{Braguta:2023yjn}. Note that negative angular momenta were also obtained in Ref.~\cite{Yamamoto:2013zwa} using Wilson-Dirac fermions with quenched approximation.

In the explored temperature region, $T\approx100$-$300$ MeV, $\bar\xi_{l,s}$ are found almost unchanged for different $\Omega_I$ and $T$: $\bar\xi_l=-0.0063(1) a^{-2}$ and $\bar\xi _s=-0.0068(1) a^{-2}$. Using $i\gamma _4^E \Sigma_3^{E}/2=\gamma _3^E\gamma _5^E/2$, this translates to the CVE for axial currents $J^z_{5f}=\langle\bar{q}_{f}\gamma^3\gamma_5 q_f\rangle$ under real rotation $\Omega$, $\bar J^a_{5f}=2\bar\xi_f\Omega$, with $2\bar\xi_f\approx -0.472(5) T^2$ (insensitive to flavor) re-interpreted as the (unrenormalized) CVE conductivities.

{\bf Polyakov loop and chiral condensate. ---} To investigate the rotational effects on QCD phase transitions, we measure the Polyakov loop and chiral condensate. We use the renormalized Polyakov loop~\cite{Cheng:2007jq,Bazavov:2011nk}
\begin{equation}
\begin{split}
L_{\rm ren}=\exp (-N_{\tau}c(\beta)a/2)L_{\rm bare},
\end{split}
\label{eq.4.0}
\end{equation}
where $L_{\rm bare}$ is the bare Polyakov loop defined as $L_{\rm bare}= \left|{\rm tr}\left[\sum _{{\bf n}}\prod _{\tau}U_{\tau}({\bf n}, \tau)\right]\right|/3N_x^3$, where $U_{\tau}$ is the gauge link along $\tau$ direction and $c(\beta)$ is a subtraction constant to match the static quark potential $V(r)=12\pi / r -\sigma r$ ($\sigma$ is the string tension) at $r=1.5r_0$, where $r_0=0.5\;{\rm fm}$ is the Sommer scale~\cite{Cheng:2007jq,Cheng:2009zi}. We use the renormalized chiral condensates
\begin{equation}
\begin{split}
&\Delta _{l,s}(T,\Omega_I)=\frac{\langle \bar{\psi}_l\psi_l\rangle _{T,\Omega_I} -\frac{m_l}{m_s}\langle \bar{\psi}_s\psi_s \rangle _{T,0}}{\langle \bar{\psi}_l\psi_l\rangle _{0,0}-\frac{m_l}{m_s}\langle \bar{\psi}_s\psi_s\rangle _{0,0}},
\end{split}
\label{eq.4.1}
\end{equation}
where $\langle \bar{\psi} _f\psi _f\rangle _{T,\Omega_I}$ is measured at temperature $T$ and imaginary angular velocity $\Omega_I$.
The subtracted term in the numerator is to eliminate the quadratic divergence proportional to $m_f$, and the denominator is a normalization to eliminate multiplicative renormalization factors, so that Eq.~(\ref{eq.4.1}) is finite and well defined in the continuous limit~\cite{Cheng:2007jq,Bazavov:2011nk}.

\begin{figure}[!htbp]
\begin{center}
\includegraphics[width=0.35\textwidth]{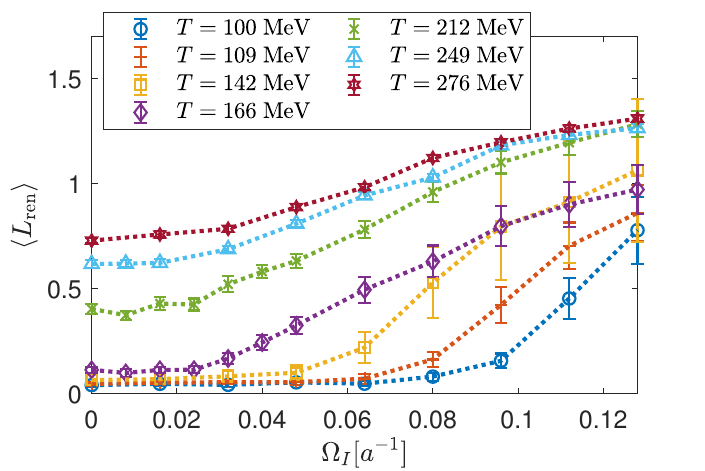}
\includegraphics[width=0.35\textwidth]{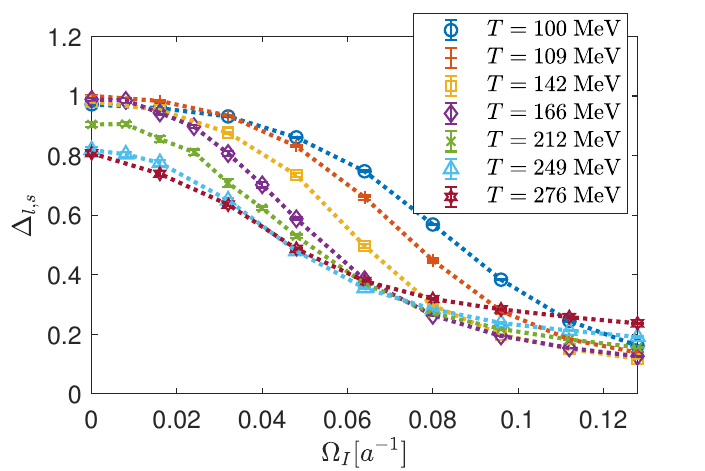}
\caption{\label{fig:polyakovandchiral}The Polyakov loop and chiral condensate as functions of $\Omega_I$.}
\end{center}
\end{figure}

In Fig.~\ref{fig:polyakovandchiral}, we show $\langle L_{\rm ren}\rangle$ and $\Delta _{l,s}$ as functions of $\Omega_I$. 
In the explored temperature region, we observe that $\langle L_{\rm ren}\rangle$ increases with $\Omega_I$ while $\Delta _{l,s}$ decreases with $\Omega_I$. This is more clearly seen at low temperatures, while at higher temperatures, both $\langle L_{\rm ren}\rangle$ and $\Delta _{l,s}$ become less sensitive to $\Omega_I$. Such behavior indicates that the imaginary rotation tends to melt the chiral condensate and to break the confinement. Similar behavior in simulations for pure gluons~\cite{Braguta:2020biu,Braguta:2021jgn} and with Wilson fermions~\cite{Braguta:2022str}.

\begin{figure}[!htbp]
\begin{center}
\includegraphics[width=0.4\textwidth]{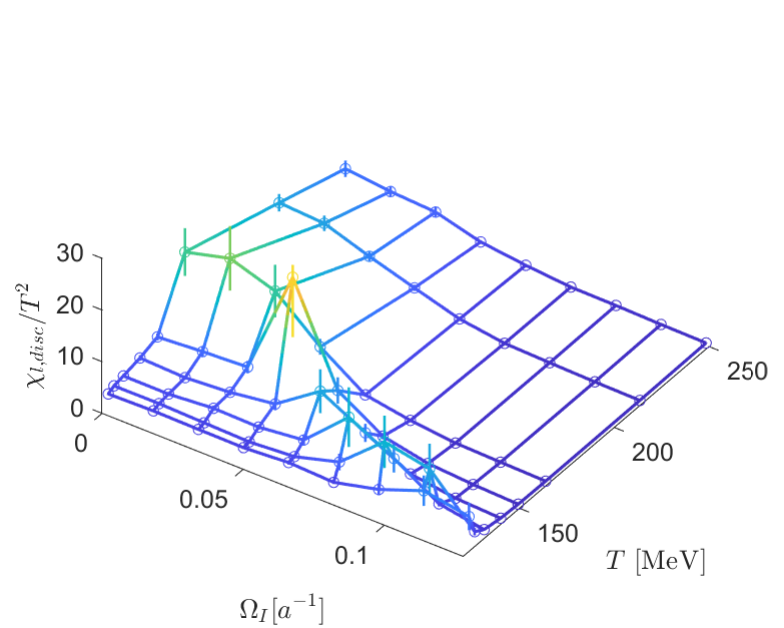}
\includegraphics[width=0.4\textwidth]{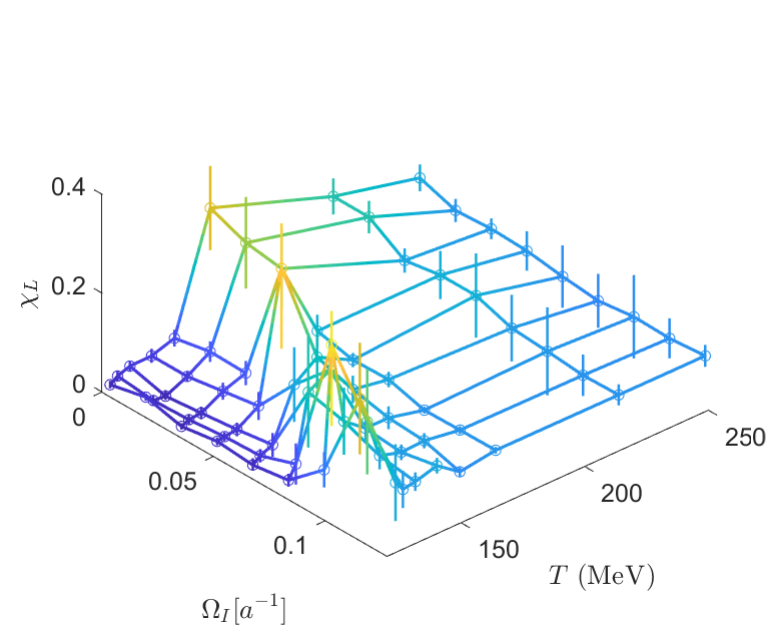}
\end{center}
\caption{\label{fig:susp}The susceptibilities $\chi _{l,\rm disc}$ and $\chi _L$ on $T-\Omega_I$ plane. The peaks are identified as the chiral and confinement-deconfinement phase boundaries, respectively.}
\end{figure}

To locate the phase transition lines, we examine the disconnected susceptibilities of chiral condensate and Polyakov loop. They are defined as $\chi _{f,{\rm disc}}=N_f^2\big[\langle {\rm tr}\big(D_f^{-1}\big)^2\rangle-\langle {\rm tr}\big(D_f^{-1}\big)\rangle^2\big]/16N_x^3N_{\tau}$ and $\chi _L=N_x^3\big(\langle L_{\rm bare}\rangle ^2 -\langle L_{\rm bare}^2\rangle\big)$~\cite{Bazavov:2011nk}. The critical imaginary angular velocities, $\Omega_{Ic}$'s, for chiral and confinement-deconfinement phase transitions are determined according to the peaks of $\chi _{l,\rm disc}$ and $\chi_L$, respectively, which are shown in Fig.~\ref{fig:susp} with $N_{\tau}=4$ being used. It can be found that, $\Omega _{Ic}$'s for chiral and confinement-deconfinement phase transitions almost coincide with each other and they both decrease with decreasing temperature, exhibiting the (imaginary) rotational suppression of the critical temperatures. 

{\bf Real rotation. ---} Since an imaginary angular velocity $\Omega_I$ is used in the simulation, an analytical continuation is needed to obtain the corresponding results for real rotation $\Omega$. For an observable $O(\Omega)$ analytical in a domain $|\Omega|<1/R$ ($R$ is the transverse radius of the system) on the complex $\Omega$ plane, this is achieved by the replacement $\Omega_I\to -i\Omega$. For a moment, let us suppose that the chiral condensate and the Polyakov loop are such observable. Given that both the chiral condensate and the Polyakov loop are even functions of $\Omega_I$, our simulations suggest that a real uniform rotation enhances the chiral condensate and suppresses the Polyakov loop; that is, we find a {\it rotational catalysis} of chiral symmetry breaking and confinement of QCD by real uniform rotation at finite temperatures. The result about the chiral condensate sharply contradicts the studies based on effective models which predict a suppression of the chiral condensate by real rotation~\cite{Chen:2015hfc,Jiang:2016wvv,Chernodub:2016kxh,Chernodub:2017ref,Wang:2018sur,Wang:2018zrn,Wang:2019nhd,Zhang:2020jux,Sadooghi:2021upd,Chen:2022mhf,Chen:2023cjt}. One possible reason is that these effective models do not include properly the contribution from gluon dynamics. A recent study based on Nambu-Jona-Lasinio model showed that, once a gluon-dressed four-fermion coupling is introduced, the enhancement of chiral condensate by real rotation can occur~\cite{Jiang:2021izj}. Other non-perturbative tools, e.g., the functional renormalization group and Dyson-Schwinger equation method, may provide a test for this.

Our result about the Polyakov loop extends and supports the previous results for pure $SU(3)$ gluondynamics~\cite{Braguta:2020biu,Braguta:2021jgn}, but contradicts with previous model studies~\cite{Chen:2020ath,Braga:2022yfe,Yadav:2022qcl,Zhao:2022uxc,Fujimoto:2021xix,Chen:2022smf} which predict a catalysis of deconfinement by real rotation. Such a contradiction makes some of these works question the validation of analytical continuation around $\Omega\sim 0$~\cite{Chen:2022smf,Chernodub:2022veq,Chernodub:2022qlz}~\footnote{In these studies, a finite imaginary rotation is introduced by imposing a twisted boundary condition in the imaginary-time direction (e.g., for gluons, $A^a_\mu(\tau, r, \phi, z)=A^a_\mu(\tau+1/T, r, \phi-\Omega_I/T, z)$) rather than by going into the rotating frame as we used.}. To test the validation of analytical continuation, we consider the Polyakov loop at quenched limit. For a small real rotation, we have the Taylor expansion $\langle L_{\rm bare}\rangle(\Omega)=c_0+c_2(a\Omega)^2 + \cdots$ (The odd powers vanish because $\langle L_{\rm bare}\rangle(\Omega)$ is time-reversal even). Choosing lattice coupling $\beta=5.7$ and $N_x^3\times N_{\tau}=12^3\times 4$ as an example, we obtain $c_0=0.10498(3)$ and $c_2=(-3.0\pm 1.1)\times 10^2$, which indeed shows that a real uniform rotation suppresses the Polyakov loop. A detail comparison between the Taylor expansion and analytical continuation $\Omega_I\to -i\Omega$ is given in the Appendix.

{\bf Summary. ---} In this paper, we studied rotating hot QCD matter using lattice approach. The rotation was implemented as simulating a rest state in a rotating frame with an imaginary angular velocity and with a special periodic boundary condition such that the ${xy}$ plane is a projective plane. We computed different components of QCD angular momentum and obtained the rotational rigidity (see definition in the main text) and spin susceptibility. We observed a negative QCD rotational rigidity which may indicate a possible thermodynamic instability against uniform rotation. We also computed the Polyakov loop and chiral condensate as well as the corresponding susceptibilities. We found that the imaginary rotation suppresses both confinement-deconfinement and chiral critical temperatures. This translates to a surprising conclusion after analytic continuation to real rotation: the real uniform rotation tends to catalyze the chiral symmetry breaking and color confinement. This conflicts sharply with previous model studies. Further theoretical and numerical works are necessary to fully understand the underlying mechanism. 

{\bf Acknowledgement.---} We thank H.-L. Chen, K. Fukushima, K. Mameda, A. Yamamoto for useful discussions. X.-G.H. is supported by the  Natural Science Foundation of China (Grant No. 12147101, No. 12225502 and No. 12075061), the National Key Research and Development Program of China (Grant No. 2022YFA1604900), and the Natural Science Foundation of Shanghai (Grant No. 20ZR1404100). J.-C. Y. is supported in part by the Natural Science Foundation of China~(No.~12147214), the Natural Science Foundation of the Liaoning Scientific Committee~(No.~LJKZ0978).

\appendix
\section{Appendix}

\begin{figure*}[!htbp]
    \begin{center}
    \includegraphics[width=0.78\textwidth]{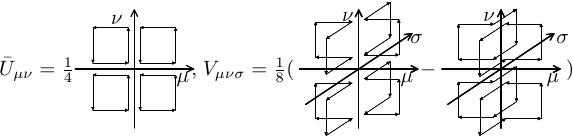}
    \caption{\label{fig:uandv}Graphical representations of $\bar{U}_{\mu\nu}$ and $V_{\mu\nu\sigma}$.}
    \end{center}
\end{figure*}

{\bf Lattice discretization. ---} The rotation is along $z$-axis. We discretize the gauge action $S_{\rm G}$ as
\begin{equation}
\begin{split}
&S_{\rm G}=\frac{\beta}{N_c}\sum _{n}\Big\{\sum _{\mu >\nu}{\rm Re\,tr}[1-\bar{U}_{\mu\nu}(n)]-\Omega_I\big(x{\rm Re\,tr}[V_{\tau xy}(n)\\
&\qquad+V_{\tau zy}(n)]-y{\rm Re\,tr}[V_{\tau yx}(n)+V_{\tau zx}(n)]\big)\\
&\qquad+\Omega_I^2\left(x^2{\rm Re\,tr}[1-\bar{U}_{yz}(n)]+y^2{\rm Re\,tr}[1-\bar{U}_{xz}(n)]\right.\\
&\left.\qquad+(x^2+y^2){\rm Re\,tr}[1-\bar{U}_{xy}(n)]-xy{\rm Re\,tr}[V_{xzy}(n)]\right)\Big\},
\end{split}
\label{eq.gaugeaction}
\end{equation}
where $\bar{U}_{\mu\nu}$ and $V_{\mu\nu\sigma}$ are defined in Ref.~\cite{Yamamoto:2013zwa} and depicted in Fig.~\ref{fig:uandv}. 

The fermion action is given in Eq.~(\ref{eq.2.4}). The angular momentum density (only $z$ component is nonzero) is discretized in the same way as the action. Under imaginary rotation, the angular momentum is also imaginary. Thus we measure the imaginary part of the angular momentum on lattice which, after analytical continuation to real rotation, gives the angular momentum under real rotation. The results for the different components of the angular momentum are
\begin{equation}
\begin{split}
J^z_{\rm G}(n)&=-\frac{a^{-4}\beta}{N_c}\big\{x{\rm Re\, tr}[V_{\tau x y}(n)+V_{\tau z y}(n)]\\&\quad-y{\rm Re\, tr}[V_{\tau y x}(n)+V_{\tau z x}](n)\big\},\\
L^z_f(n) &=\frac{a^{-4}}{4}\left(b_{x,y}-b_{y,x}\right),\\
s^z_f(n) &=\frac{a^{-3}}{8}\sum _{s_{x,y,\tau}=\pm 1}s_{\tau}\eta _{s_{\tau}\tau}(n)\eta _{xy}(n) \bar{\psi} (n)\\
&\quad\times U\Big(n,n+\sum _{i=x,y,\tau}s_i \hat{i}\Big)\psi \Big(n+\sum _{i=x,y,\tau}s_i \hat{i}\Big).
\end{split}
\label{eq.a.9}
\end{equation}
The rotational rigidities and the spin susceptibility are averaged on lattice as
\begin{equation}
    \begin{split}
    &\bar\xi _f = \frac{1}{N_{\rm taste}N_{r_{\rm max}}}\sum _{n_x^2+n_y^2<r_{\rm max}^2}\frac{ s^z_f(n) }{\Omega_I},\\
    &\bar\rho _{J_G}= \frac{1}{N_{r_{\rm max}}}\sum _{n_x^2+n_y^2<r_{\rm max}^2}\frac{ J^z_G(n) }{\Omega_I r^2},
 \end{split}
    \label{eq.4.6}
    \end{equation}
and similarly for $\bar\rho_{L_f}$. Here, $N_{r_{\rm max}}$ is the number of sites satisfying $n_x^2+n_y^2<r_{\rm max}^2$, $N_{\rm taste}$ is the taste degeneracy ($N_{\rm taste}=4$ for $L_f$). We choose $r_{\rm max}=6, 5$, and $7$ for $J_G$, $L_f$, and $s_f$ in the simulation, respectively.

{\bf Equivalence with rotating ensemble in inertial frame. ---} In the main text, we have addressed that the rotation is implemented by simulating a rest state in a rotating frame. Here, we demonstrate the equivalence of this approach with simulating a rotating state in an inertial frame using the transfer matrix method~\cite{Fradkin:1978th,Kogut:1979wt}. For the sake of clarity and simplicity, we consider a pure $U(1)$ gauge theory as an example. In this case, Eq.~(\ref{eq.gaugeaction}) can be written as,
\begin{equation}
\begin{split}
&S_G=a_s^3 a_{\tau}\sum _{n}\left\{\sum _{\mu >\nu}\frac{\theta _{\mu\nu}^2}{2}\right.\\
&\left.+\Omega _I\left[x\left(\theta _{\tau x}(n)\theta _{xy}(n)+\theta _{\tau z}(n)\theta _{zy}(n)\right)\right.\right.\\
&\left.\left.-y\left(\theta_{\tau y}(n)\theta _{yx}(n)+\theta_{\tau z}(n)\theta _{zx}(n)\right)\right]\right.\\
&\left.+\frac{\Omega _I^2}{2}\left[\left(x\theta _{zy}(n)-y\theta _{zx}(n)\right)^2+r^2\theta ^2_{xy}(n)\right]\right\},\\
\end{split}
\label{eq.actionimageu1}
\end{equation}
where $\theta _{\mu\nu}({n})\equiv \left(\Delta _{\mu}\theta _{\nu}(n)/a_{\mu}\right)-\left(\Delta _{\nu}\theta _{\mu}(n)/a_{\nu}\right)$ (To avoid confusion, we use $\theta _{\mu\nu}$ instead of $F_{\mu\nu}$ to denote $U(1)$ gauge-field strength tensor), and $\Delta _{\mu}\theta _{\nu}(n)\equiv \theta _{\nu}(n)-\theta _{\nu}(n-\mu)$, $r^2=x^2+y^2$, $a_s$ and $a_{\tau}$ are lattice spacings of spatial and time directions, respectively.
In temporal gauge, $\theta _{\tau}=0, \theta _{\tau i}=\Delta _{\tau} \theta _i(n)/a_{\tau}$, and we have
\begin{equation}
\begin{split}
&S_G=\sum _{n}\left\{\frac{1}{2}\sum _{i >j}\theta _{ij}^2(n)+\frac{1}{2}\left(\frac{\Delta _{\tau}}{a_{\tau}} \theta _x(n)+x\Omega _I\theta _{xy}(n)\right)^2\right.\\
&\left.+\frac{1}{2}\left(\frac{\Delta _{\tau}}{a_{\tau}} \theta _y(n)+y\Omega _I\theta _{xy}(n)\right)^2\right.\\
&\left.+\frac{1}{2}\left(\frac{\Delta _{\tau}}{a_{\tau}} \theta _z(n)+x\Omega _I\theta _{zy}(n)-y\Omega _I\theta _{zx}(n)\right)^2\right\}.\\
\end{split}
\label{eq.actionimageu1temporal}
\end{equation}
Equation~(\ref{eq.actionimageu1temporal}) is a sum of actions which depend on only two neighboring time slices, and therefore the partition function is $Z=\sum_{\{\theta\}}\exp (-S_G)=\sum_{\{\theta\}}\prod _\tau T(\tau+1,\tau)$ with
\begin{equation}
\begin{split}
&-\log T(\tau+1,\tau)=a_{\tau}a_s^3\sum _{{\bf n}}\left\{\frac{1}{4}\sum _{i >j}\theta _{ij}^2+ \frac{1}{4}\sum _{i >j}{\theta'} _{ij}^2\right.\\
&\left.+\frac{1}{2a_{\tau}^2} \left[
\left(\theta'_x-\theta _x+x\Omega _I a_{\tau}\theta _{xy}\right)^2 +\left(\theta'_y-\theta _y+y\Omega _I a_{\tau}\theta _{xy}\right)^2\right] \right.\\
&\left.+\frac{1}{2a_{\tau}^2}\left(\theta'_z-\theta _z+\Omega _I a_{\tau}\left(x\theta _{zy}-y\theta _{zx}\right)\right)^2 \right\},\\
\end{split}
\end{equation}
where $\theta ' ({\bf n})= \theta({\bf n}, \tau+1), \theta({\bf n}) = \theta({\bf n},\tau)$.

Introducing generalized coordinate operator $\hat{\theta}({\bf n})$ and corresponding generalized momentum operator $\hat{L}({\bf n})$ satisfying $[\hat{L} _i({\bf n}),\hat{\theta} _j({\bf n'})]=-i\delta _{ij}\delta _{{\bf n},{\bf n'}}$, and using
\begin{equation}
\begin{split}
&\langle \theta' | \exp (-\frac{1}{2}\frac{a_{\tau}}{a_s^3} \hat{L}^2 )\exp(ia_{\tau}\hat{L}f(\hat{\theta}))|\theta\rangle  \\
&={\rm const.}\times \exp \left(-\frac{a_s^3}{2a_{\tau}}\left[\theta'-\theta+a_{\tau}f(\theta)\right]^2\right)+\mathcal{O}(a_{\tau}^2)\\
\end{split}
\end{equation}
where $f(\hat{\theta})$ is an arbitrary function of coordinate operator only, $T(\tau+1,\tau)$ can be written as an operator sandwiched with $\{\theta'\}$ and $\{\theta\}$ configurations,
\begin{equation}
\begin{split}
&T(\tau,\tau+1)=\langle \theta' |\exp\left\{-a_s^3a_{\tau}\frac{1}{4}\sum _{i >j}\hat{\theta} _{ij}^2\right\}\\
&\times  \exp\left\{-\frac{a_{\tau}}{2a_s^3}\sum _i\hat{L}_i^2 + ixa_{\tau}\Omega _I\hat{L}_x\hat{\theta} _{xy}+ iya_{\tau}\Omega _I\hat{L}_y\hat{\theta} _{xy}\right.\\
&\left. +ia_{\tau}\Omega _I\hat{L}_z\left(x\hat{\theta} _{zy}-y\hat{\theta} _{zx}\right) \right\}\\
&\times  \exp\left\{-a_s^3a_{\tau}\frac{1}{4}\sum _{i >j}\hat{\theta} _{ij}^2\right\}|\theta \rangle + \mathcal{O}(a_{\tau}^2).\\
\end{split}
\end{equation}
As a result, $T(\tau+1,\tau)=\langle \theta '|\hat{T}|\theta\rangle + \mathcal{O}(a_{\tau}^2)$ with,
\begin{equation}
\begin{split}
&\hat{T}=\exp \left\{a_{\tau}\left(-\frac{1}{2a_s^3}\sum _i\hat{L}_i^2 -a_s^3\frac{1}{2}\sum _{i >j}\hat{\theta} _{ij}^2\right.\right.\\
&\left.\left.+ ix\Omega _I\hat{L}_x\hat{\theta} _{xy}+ iy\Omega _I\hat{L}_y\hat{\theta} _{xy} +i\Omega _I\hat{L}_z\left(x\hat{\theta} _{zy}-y\hat{\theta} _{zx}\right) \right)\right\}.
\end{split}
\end{equation}
Therefore, the partition function can be written as
\begin{equation}
\begin{split}
&Z=\sum _{\{\theta ^1,\theta ^2,\ldots \}}\langle \theta ^{N_{\tau}} |\hat{T}|\theta ^{N_{\tau}-1} \rangle \ldots \langle \theta ^3 | \hat{T}|\theta ^2 \rangle \langle \theta ^2 | \hat{T}|\theta ^1 \rangle ,\\
\end{split}
\end{equation}
where $\{\theta^i\}$ is the configuration at $\tau_i$ time slice.
By requiring periodic boundary condition in the time direction $\theta ^1=\theta ^{N_{\tau}}$, $Z={\rm tr}\left[\hat{T}^{N_{\tau}}\right]$. Compared with $Z={\rm tr}\left[\exp \left(- \hat{H}/T\right)\right]$, the Hamiltonian operator can be read out as
\begin{equation}
\begin{split}
&\hat{H}=\sum _{\bf n}\left(\frac{1}{2a_s^3}\sum _i\hat{L}_i^2 +a_s^3\frac{1}{2}\sum _{i >j}\hat{\theta} _{ij}^2\right.\\
&\left.- i\Omega _I\left( (x \hat{L}_x + y \hat{L}_y)\hat{\theta} _{xy}+ \hat{L}_z\left(x\hat{\theta} _{zy}-y\hat{\theta} _{zx}\right) \right)\right),\\
\end{split}
\label{eq.hamiltoniangauge}
\end{equation}
with $T = a_{\tau}^{-1}/N_{\tau}$. Equation~(\ref{eq.hamiltoniangauge}) is nothing but the Hamiltonian for an rotating $U(1)$ system in inertial frame, $\hat{H}=\hat{H}_0+i\Omega _I \hat{J}^z_G$, with $\hat{H}_0$ the Hamiltonian without rotation and
\begin{equation}
\begin{split}
&\hat{J}^z_G = - \left( (x \hat{L}_x + y \hat{L}_y)\hat{\theta} _{xy}+ \hat{L}_z\left(x\hat{\theta} _{zy}-y\hat{\theta} _{zx}\right) \right).\\
\end{split}
\label{eq.jzg}
\end{equation}
being just the angular momentum operator. This shows that our lattice action~[Eq.~(\ref{eq.actionimageu1})], introduced by considering a rest state in rotating frame, represents a rotating state in an inertial frame.

{\bf Spinor eigenstates for projective plane boundary condition. ---} In cylindrical coordinate, the general spinor eigenstates can be written as~\cite{Jiang:2016wvv}
\begin{equation}
\begin{split}
&u=\mathcal{N}_k e^{in\theta+ik_zz}\begin{pmatrix}
J_n(k_tr)\\
se^{i\theta}J_{n+1}(k_tr)\\
\frac{k_z-isk_t}{E_k+m}J_n(k_tr)\\
\frac{ik_t-sk_z}{E_k+m}e^{i\theta}J_{n+1}(k_tr)\\ \end{pmatrix},\\
&v= \mathcal{N}_k e^{in\theta-ik_zz}\begin{pmatrix}
\frac{k_z-isk_t}{E_k+m}J_n(k_tr)\\
\frac{ik_t-sk_z}{E_k+m}e^{i\theta}J_{n+1}(k_tr)\\ J_n(k_tr)\\
-se^{i\theta}J_{n+1}(k_tr)\\ \end{pmatrix},\\
\end{split}
\label{eq.a.8}
\end{equation}
for particle and anti-particle modes, respectively, where $k_{z,t}$ are momenta along and transverse to $z$-axis, $n\in \mathbb{Z}$, $s=\pm$ is the transverse helicity, $J_n(x)$ is the Bessel function of the first kind, and ${\mathcal N}_k$ is a normalization factor.

The Dirichlet boundary condition, $u(R)=v(R)=0$, is not feasible because the zeros of $J_n(x)$ and $J_{n+1}(x)$ are different. However, noting that functions with the form $f(\theta, r)=e^{in \theta} g(r)$ satisfy the projective plane boundary condition when $n$ is even.
Therefore, for an even (odd) $n$, one can choose $k_t$ such that $J_{n+1}(k_tR)=0$ [$J_{n}(k_tR)=0$]. In this way, we find that the spinor eigenstates $u$ and $v$ are compatible with the projective plane boundary condition.

{\bf Lattice parameters. ---} The lattice spacing is matched by measuring the static quark potential $V(r)$~\cite{Cheng:2007jq} and Sommer scale $r_{0}$~\cite{Sommer:1993ce} at low temperature and zero angular velocity with projective plane boundary condition using the methods described in Refs.~\cite{Bali:1992ab,Bali:2000vr,Orth:2005kq}. Adopting $r_0=0.5\;{\rm fm}$~\cite{Sommer:1993ce}, the lattice coupling $\beta$ and the corresponding lattice spacings are listed in Table~\ref{tab:matching1}. Denoting the molecular-dynamics time unit as $TU$, when doing the matching, $TU_{\rm th}$ trajectories are discarded for thermalization, and $TU_{m}$ configurations are measured. Throughout this work, the statistical error is estimated as $\sigma =\sqrt{2TU _{\rm int}}\sigma _{jk}$ where $\sigma _{jk}$ is calculated by jackknife method, $2TU _{\rm int}$ is the span of $TU$ when the two configurations can be regarded as being independent, and $TU _{\rm int}$ is calculated by using the autocorrelation method with $S=1.5$~\cite{Wolff:2003sm} on the bare chiral condensate of light quarks defined as $\langle \bar{\psi} _l\psi _l\rangle  = -{\rm tr}\left[D_l^{-1}\right] / 4 N_x^3N_{\tau}$. 
\begin{table}[!htbp]
\begin{tabular}{c|c|c|c|c|c}
$\beta$  & $am_l$ & $r_0/a$ & $a^{-1}\;({\rm MeV})$ & $c(\beta)r_0$  & $\langle \bar{\psi}\psi\rangle_l -\frac{m_l}{m_s}\langle \bar{\psi}\psi\rangle _s$ \\
\hline
$4.72$ & $0.045$   & $1.250(28)$  & $493(11)$  & $-0.79(29)$ & $0.1263(1)$ \\
$4.75$ & $0.04$    & $1.280(24)$  & $505(10)$  & $-0.94(28)$ & $0.1231(1)$ \\
$4.78$ & $0.036$   & $1.330(17)$  & $525(7)$   & $-1.17(25)$ & $0.1198(1)$ \\
$4.82$ & $0.0325$  & $1.419(8)$   & $560(3)$   & $-1.19(14)$ & $0.1152(1)$ \\
$4.86$ & $0.0295$  & $1.514(11)$  & $597(5)$   & $-1.27(10)$ & $0.1129(1)$ \\
$4.90$ & $0.027$   & $1.661(17)$  & $655(7)$   & $-1.38(9)$  & $0.1031(1)$ \\
$4.94$ & $0.0235$  & $2.163(81)$  & $854(32)$  & $-1.86(24)$ & $0.0944(2)$ \\
$4.98$ & $0.02$    & $2.519(54)$  & $994(21)$  & $-2.02(11)$ & $0.0769(2)$ \\
$5.02$ & $0.0165$  & $3.220(56)$  & $1271(22)$ & $-2.78(7)$  & $0.0598(2)$ \\
$5.06$ & $0.013$   & $3.779(29)$  & $1491(11)$ & $-3.19(3)$  & $0.0467(3)$ \\
$5.10$ & $0.011$   & $4.201(34)$  & $1658(14)$ & $-3.49(3)$  & $0.0364(1)$ \\
$5.70$ & -  & $3.006(5)$  & $1186(2)$ & -  & - \\
\hline
\end{tabular}
\caption{\label{tab:matching1}The simulation parameters and the lattice spacing matched by using $r_0=0.5\;{\rm fm}$ at zero T and zero rotation. The normalization parameters $c(\beta)$ and $\langle \bar{\psi}\psi\rangle_l -\frac{m_l}{m_s}\langle \bar{\psi}\psi\rangle _s$ at zero T and zero rotation are also shown.}
\end{table}

When the rotation is turned on, for each $\beta$, $1.5TU_{\rm th}+(TU_{\rm th}+TU_{m})\times (K+1)$ trajectories with $K=8$ are simulated sequentially with increasing $\Omega_I=\Delta \Omega_I \times k$, $0\leq k\leq K$. Here, in the case of $N_{\tau}=6$, $TU_{\rm th}=100$, $TU_{m}=1900, 3400$ for $\beta=4.98,5.1$, respectively, and $TU_m=1400$ for other $\beta$'s in Table~\ref{tab:matching1} and in $\beta=4.86\sim 5.1$, and in the case of $N_{\tau}=4$, $TU_{\rm th}=100$, $TU_{m}=900$ for $\beta=4.72\sim 4.98$, and $\Delta \Omega_I = \Omega _{I\rm max} / K$ with $\Omega _{I\rm max}$ approximately the maximally allowed angular velocity on the lattice when Wick transformed into real rotation. Therefore, with $N_x=12$, $a\Omega _{I\rm max} = 0.128$ is used such that the maximal linear velocity is $a\Omega _{I\rm max} \times 11\sqrt{2}/2 \approx 0.996<1$.

\begin{figure}[!htbp]
    \begin{center}
    \includegraphics[width=0.35\textwidth]{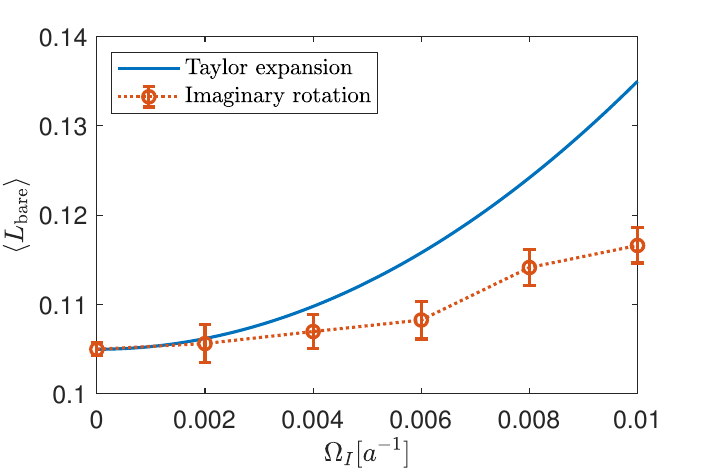}\\
    \end{center}
    \caption{\label{fig:taylor} Polyakov loop at small rotation by Taylor expansion with real rotation followed by an analytical continuation and by a direct simulation with imaginary rotation.}
\end{figure}

{\bf Small real angular velocity and Taylor expansion. ---} Consider a real rotation with angular velocity $\Omega$ along $z$-axis. The action is expanded as $S=S_{0}+a\Omega S_{\Omega }+(a\Omega)^2 S_{\Omega^2} + \cdots$. For an operator $O$ that is time-reversal even and does not depend on $\Omega$ explicitly (the operators for chiral condensate and Polyakov loop are both such operators), we have the following Taylor expansion for $\langle O\rangle$,
\begin{equation}
\langle O\rangle(\Omega)= c_0+c_2(a\Omega)^2 + \mathcal{O}(a\Omega)^4.
\label{apptaylor}
\end{equation}
The coefficients $c_0=\langle O\rangle_0$ and
\begin{equation}
\begin{split}
c_2 &=\frac{1}{2}\frac{d^2\langle O\rangle}{d(a\Omega)^2}\Big|_{\Omega=0}\\
&= \frac{1}{2}\langle O\rangle _0 \langle (2S_{\Omega^2}-S_{\Omega}^2)\rangle _0 - \frac{1}{2}\langle O (2S_{\Omega^2}-S_{\Omega}^2) \rangle _0 ,\
\end{split}
\label{apptaylor2}
\end{equation}
where $\langle \ldots \rangle _0$ denotes average at $\Omega = 0$. For $O=L_{\rm ren}$, using quenched approximation at lattice coupling $\beta = 5.7$ on a $N_x^3\times N_\tau=12^3\times 4$ lattice (temperature $T=296.5(5)$ MeV) with the projective plane boundary condition, $3\times 10^{6}$ configurations are generated, and we find
\begin{equation}
\begin{split}
\langle L_{\rm bare} \rangle&=0.10498(3)+ \left(-3.0\pm 1.1\right) \times 10^2(a\Omega)^2+ \mathcal{O}(a\Omega)^4,\\
\end{split}
\end{equation}
which supports the conclusion that the real rotation drives the rotating hot QCD towards the confinement phase. Supposing that an analytical continuation to imaginary rotation by replacing $\Omega$ with $i\Omega_I$ is allowed, we then make a comparison with the simulation directly performed for imaginary rotation, see Fig.~\ref{fig:taylor}. It can be seen that, $\partial ^2 L_{\rm ren} / \partial (a\Omega) ^2 \approx -10^3 $, which implies that the Taylor expansion is poorly converged, and the result of Taylor expansion is reliable only when $a\Omega \ll 1/\sqrt{10^3}$, which explains the deviation of analytical continuation from the result of direct simulation with imaginary rotation in Fig.~\ref{fig:taylor}. Nevertheless, the trends and orders of magnitude are consistent between the two methods.

\bibliography{RotatingStaggered}

\begin{thebibliography}{58}%
\makeatletter
\providecommand \@ifxundefined [1]{%
 \@ifx{#1\undefined}
}%
\providecommand \@ifnum [1]{%
 \ifnum #1\expandafter \@firstoftwo
 \else \expandafter \@secondoftwo
 \fi
}%
\providecommand \@ifx [1]{%
 \ifx #1\expandafter \@firstoftwo
 \else \expandafter \@secondoftwo
 \fi
}%
\providecommand \natexlab [1]{#1}%
\providecommand \enquote  [1]{``#1''}%
\providecommand \bibnamefont  [1]{#1}%
\providecommand \bibfnamefont [1]{#1}%
\providecommand \citenamefont [1]{#1}%
\providecommand \href@noop [0]{\@secondoftwo}%
\providecommand \href [0]{\begingroup \@sanitize@url \@href}%
\providecommand \@href[1]{\@@startlink{#1}\@@href}%
\providecommand \@@href[1]{\endgroup#1\@@endlink}%
\providecommand \@sanitize@url [0]{\catcode `\\12\catcode `\$12\catcode
  `\&12\catcode `\#12\catcode `\^12\catcode `\_12\catcode `\%12\relax}%
\providecommand \@@startlink[1]{}%
\providecommand \@@endlink[0]{}%
\providecommand \url  [0]{\begingroup\@sanitize@url \@url }%
\providecommand \@url [1]{\endgroup\@href {#1}{\urlprefix }}%
\providecommand \urlprefix  [0]{URL }%
\providecommand \Eprint [0]{\href }%
\providecommand \doibase [0]{https://doi.org/}%
\providecommand \selectlanguage [0]{\@gobble}%
\providecommand \bibinfo  [0]{\@secondoftwo}%
\providecommand \bibfield  [0]{\@secondoftwo}%
\providecommand \translation [1]{[#1]}%
\providecommand \BibitemOpen [0]{}%
\providecommand \bibitemStop [0]{}%
\providecommand \bibitemNoStop [0]{.\EOS\space}%
\providecommand \EOS [0]{\spacefactor3000\relax}%
\providecommand \BibitemShut  [1]{\csname bibitem#1\endcsname}%
\let\auto@bib@innerbib\@empty
\bibitem [{\citenamefont {Paschalidis}\ and\ \citenamefont
  {Stergioulas}(2017)}]{Paschalidis:2016vmz}%
  \BibitemOpen
  \bibfield  {author} {\bibinfo {author} {\bibfnamefont {V.}~\bibnamefont
  {Paschalidis}}\ and\ \bibinfo {author} {\bibfnamefont {N.}~\bibnamefont
  {Stergioulas}},\ }\bibfield  {title} {\bibinfo {title} {{Rotating Stars in
  Relativity}},\ }\href {https://doi.org/10.1007/s41114-017-0008-x} {\bibfield
  {journal} {\bibinfo  {journal} {Living Rev. Rel.}\ }\textbf {\bibinfo
  {volume} {20}},\ \bibinfo {pages} {7} (\bibinfo {year} {2017})},\ \Eprint
  {https://arxiv.org/abs/1612.03050} {arXiv:1612.03050 [astro-ph.HE]}
  \BibitemShut {NoStop}%
\bibitem [{\citenamefont {Deng}\ and\ \citenamefont
  {Huang}(2016)}]{Deng:2016gyh}%
  \BibitemOpen
  \bibfield  {author} {\bibinfo {author} {\bibfnamefont {W.-T.}\ \bibnamefont
  {Deng}}\ and\ \bibinfo {author} {\bibfnamefont {X.-G.}\ \bibnamefont
  {Huang}},\ }\bibfield  {title} {\bibinfo {title} {{Vorticity in Heavy-Ion
  Collisions}},\ }\href {https://doi.org/10.1103/PhysRevC.93.064907} {\bibfield
   {journal} {\bibinfo  {journal} {Phys. Rev. C}\ }\textbf {\bibinfo {volume}
  {93}},\ \bibinfo {pages} {064907} (\bibinfo {year} {2016})},\ \Eprint
  {https://arxiv.org/abs/1603.06117} {arXiv:1603.06117 [nucl-th]} \BibitemShut
  {NoStop}%
\bibitem [{\citenamefont {Jiang}\ \emph {et~al.}(2016)\citenamefont {Jiang},
  \citenamefont {Lin},\ and\ \citenamefont {Liao}}]{Jiang:2016woz}%
  \BibitemOpen
  \bibfield  {author} {\bibinfo {author} {\bibfnamefont {Y.}~\bibnamefont
  {Jiang}}, \bibinfo {author} {\bibfnamefont {Z.-W.}\ \bibnamefont {Lin}},\
  and\ \bibinfo {author} {\bibfnamefont {J.}~\bibnamefont {Liao}},\ }\bibfield
  {title} {\bibinfo {title} {{Rotating quark-gluon plasma in relativistic heavy
  ion collisions}},\ }\href {https://doi.org/10.1103/PhysRevC.94.044910}
  {\bibfield  {journal} {\bibinfo  {journal} {Phys. Rev. C}\ }\textbf {\bibinfo
  {volume} {94}},\ \bibinfo {pages} {044910} (\bibinfo {year} {2016})},\
  \bibinfo {note} {[Erratum: Phys.Rev.C 95, 049904 (2017)]},\ \Eprint
  {https://arxiv.org/abs/1602.06580} {arXiv:1602.06580 [hep-ph]} \BibitemShut
  {NoStop}%
\bibitem [{\citenamefont {Erdmenger}\ \emph {et~al.}(2009)\citenamefont
  {Erdmenger}, \citenamefont {Haack}, \citenamefont {Kaminski},\ and\
  \citenamefont {Yarom}}]{Erdmenger:2008rm}%
  \BibitemOpen
  \bibfield  {author} {\bibinfo {author} {\bibfnamefont {J.}~\bibnamefont
  {Erdmenger}}, \bibinfo {author} {\bibfnamefont {M.}~\bibnamefont {Haack}},
  \bibinfo {author} {\bibfnamefont {M.}~\bibnamefont {Kaminski}},\ and\
  \bibinfo {author} {\bibfnamefont {A.}~\bibnamefont {Yarom}},\ }\bibfield
  {title} {\bibinfo {title} {{Fluid dynamics of R-charged black holes}},\
  }\href {https://doi.org/10.1088/1126-6708/2009/01/055} {\bibfield  {journal}
  {\bibinfo  {journal} {JHEP}\ }\textbf {\bibinfo {volume} {01}},\ \bibinfo
  {pages} {055}},\ \Eprint {https://arxiv.org/abs/0809.2488} {arXiv:0809.2488
  [hep-th]} \BibitemShut {NoStop}%
\bibitem [{\citenamefont {Banerjee}\ \emph {et~al.}(2011)\citenamefont
  {Banerjee}, \citenamefont {Bhattacharya}, \citenamefont {Bhattacharyya},
  \citenamefont {Dutta}, \citenamefont {Loganayagam},\ and\ \citenamefont
  {Surowka}}]{Banerjee:2008th}%
  \BibitemOpen
  \bibfield  {author} {\bibinfo {author} {\bibfnamefont {N.}~\bibnamefont
  {Banerjee}}, \bibinfo {author} {\bibfnamefont {J.}~\bibnamefont
  {Bhattacharya}}, \bibinfo {author} {\bibfnamefont {S.}~\bibnamefont
  {Bhattacharyya}}, \bibinfo {author} {\bibfnamefont {S.}~\bibnamefont
  {Dutta}}, \bibinfo {author} {\bibfnamefont {R.}~\bibnamefont {Loganayagam}},\
  and\ \bibinfo {author} {\bibfnamefont {P.}~\bibnamefont {Surowka}},\
  }\bibfield  {title} {\bibinfo {title} {{Hydrodynamics from charged black
  branes}},\ }\href {https://doi.org/10.1007/JHEP01(2011)094} {\bibfield
  {journal} {\bibinfo  {journal} {JHEP}\ }\textbf {\bibinfo {volume} {01}},\
  \bibinfo {pages} {094}},\ \Eprint {https://arxiv.org/abs/0809.2596}
  {arXiv:0809.2596 [hep-th]} \BibitemShut {NoStop}%
\bibitem [{\citenamefont {Landsteiner}\ \emph {et~al.}(2011)\citenamefont
  {Landsteiner}, \citenamefont {Megias},\ and\ \citenamefont
  {Pena-Benitez}}]{Landsteiner:2011cp}%
  \BibitemOpen
  \bibfield  {author} {\bibinfo {author} {\bibfnamefont {K.}~\bibnamefont
  {Landsteiner}}, \bibinfo {author} {\bibfnamefont {E.}~\bibnamefont
  {Megias}},\ and\ \bibinfo {author} {\bibfnamefont {F.}~\bibnamefont
  {Pena-Benitez}},\ }\bibfield  {title} {\bibinfo {title} {{Gravitational
  Anomaly and Transport}},\ }\href
  {https://doi.org/10.1103/PhysRevLett.107.021601} {\bibfield  {journal}
  {\bibinfo  {journal} {Phys. Rev. Lett.}\ }\textbf {\bibinfo {volume} {107}},\
  \bibinfo {pages} {021601} (\bibinfo {year} {2011})},\ \Eprint
  {https://arxiv.org/abs/1103.5006} {arXiv:1103.5006 [hep-ph]} \BibitemShut
  {NoStop}%
\bibitem [{\citenamefont {Huang}\ and\ \citenamefont
  {Sadofyev}(2019)}]{Huang:2018aly}%
  \BibitemOpen
  \bibfield  {author} {\bibinfo {author} {\bibfnamefont {X.-G.}\ \bibnamefont
  {Huang}}\ and\ \bibinfo {author} {\bibfnamefont {A.~V.}\ \bibnamefont
  {Sadofyev}},\ }\bibfield  {title} {\bibinfo {title} {{Chiral Vortical Effect
  For An Arbitrary Spin}},\ }\href {https://doi.org/10.1007/JHEP03(2019)084}
  {\bibfield  {journal} {\bibinfo  {journal} {JHEP}\ }\textbf {\bibinfo
  {volume} {03}},\ \bibinfo {pages} {084}},\ \Eprint
  {https://arxiv.org/abs/1805.08779} {arXiv:1805.08779 [hep-th]} \BibitemShut
  {NoStop}%
\bibitem [{\citenamefont {Adamczyk}\ \emph {et~al.}(2017)\citenamefont
  {Adamczyk} \emph {et~al.}}]{STAR:2017ckg}%
  \BibitemOpen
  \bibfield  {author} {\bibinfo {author} {\bibfnamefont {L.}~\bibnamefont
  {Adamczyk}} \emph {et~al.} (\bibinfo {collaboration} {STAR}),\ }\bibfield
  {title} {\bibinfo {title} {{Global $\Lambda$ hyperon polarization in nuclear
  collisions: evidence for the most vortical fluid}},\ }\href
  {https://doi.org/10.1038/nature23004} {\bibfield  {journal} {\bibinfo
  {journal} {Nature}\ }\textbf {\bibinfo {volume} {548}},\ \bibinfo {pages}
  {62} (\bibinfo {year} {2017})},\ \Eprint {https://arxiv.org/abs/1701.06657}
  {arXiv:1701.06657 [nucl-ex]} \BibitemShut {NoStop}%
\bibitem [{\citenamefont {Abdallah}\ \emph {et~al.}(2023)\citenamefont
  {Abdallah} \emph {et~al.}}]{STAR:2022fan}%
  \BibitemOpen
  \bibfield  {author} {\bibinfo {author} {\bibfnamefont {M.~S.}\ \bibnamefont
  {Abdallah}} \emph {et~al.} (\bibinfo {collaboration} {STAR}),\ }\bibfield
  {title} {\bibinfo {title} {{Pattern of global spin alignment of
  \ensuremath{\phi} and K$^{*0}$ mesons in heavy-ion collisions}},\ }\href
  {https://doi.org/10.1038/s41586-022-05557-5} {\bibfield  {journal} {\bibinfo
  {journal} {Nature}\ }\textbf {\bibinfo {volume} {614}},\ \bibinfo {pages}
  {244} (\bibinfo {year} {2023})},\ \Eprint {https://arxiv.org/abs/2204.02302}
  {arXiv:2204.02302 [hep-ph]} \BibitemShut {NoStop}%
\bibitem [{\citenamefont {Chen}\ \emph
  {et~al.}(2021{\natexlab{a}})\citenamefont {Chen}, \citenamefont {Huang},\
  and\ \citenamefont {Liao}}]{Chen:2021aiq}%
  \BibitemOpen
  \bibfield  {author} {\bibinfo {author} {\bibfnamefont {H.-L.}\ \bibnamefont
  {Chen}}, \bibinfo {author} {\bibfnamefont {X.-G.}\ \bibnamefont {Huang}},\
  and\ \bibinfo {author} {\bibfnamefont {J.}~\bibnamefont {Liao}},\ }\bibfield
  {title} {\bibinfo {title} {{QCD phase structure under rotation}},\ }\href
  {https://doi.org/10.1007/978-3-030-71427-7_11} {\bibfield  {journal}
  {\bibinfo  {journal} {Lect. Notes Phys.}\ }\textbf {\bibinfo {volume}
  {987}},\ \bibinfo {pages} {349} (\bibinfo {year} {2021}{\natexlab{a}})},\
  \Eprint {https://arxiv.org/abs/2108.00586} {arXiv:2108.00586 [hep-ph]}
  \BibitemShut {NoStop}%
\bibitem [{\citenamefont {Chen}\ \emph {et~al.}(2016)\citenamefont {Chen},
  \citenamefont {Fukushima}, \citenamefont {Huang},\ and\ \citenamefont
  {Mameda}}]{Chen:2015hfc}%
  \BibitemOpen
  \bibfield  {author} {\bibinfo {author} {\bibfnamefont {H.-L.}\ \bibnamefont
  {Chen}}, \bibinfo {author} {\bibfnamefont {K.}~\bibnamefont {Fukushima}},
  \bibinfo {author} {\bibfnamefont {X.-G.}\ \bibnamefont {Huang}},\ and\
  \bibinfo {author} {\bibfnamefont {K.}~\bibnamefont {Mameda}},\ }\bibfield
  {title} {\bibinfo {title} {{Analogy between rotation and density for Dirac
  fermions in a magnetic field}},\ }\href
  {https://doi.org/10.1103/PhysRevD.93.104052} {\bibfield  {journal} {\bibinfo
  {journal} {Phys. Rev. D}\ }\textbf {\bibinfo {volume} {93}},\ \bibinfo
  {pages} {104052} (\bibinfo {year} {2016})},\ \Eprint
  {https://arxiv.org/abs/1512.08974} {arXiv:1512.08974 [hep-ph]} \BibitemShut
  {NoStop}%
\bibitem [{\citenamefont {Jiang}\ and\ \citenamefont
  {Liao}(2016)}]{Jiang:2016wvv}%
  \BibitemOpen
  \bibfield  {author} {\bibinfo {author} {\bibfnamefont {Y.}~\bibnamefont
  {Jiang}}\ and\ \bibinfo {author} {\bibfnamefont {J.}~\bibnamefont {Liao}},\
  }\bibfield  {title} {\bibinfo {title} {{Pairing Phase Transitions of Matter
  under Rotation}},\ }\href {https://doi.org/10.1103/PhysRevLett.117.192302}
  {\bibfield  {journal} {\bibinfo  {journal} {Phys. Rev. Lett.}\ }\textbf
  {\bibinfo {volume} {117}},\ \bibinfo {pages} {192302} (\bibinfo {year}
  {2016})},\ \Eprint {https://arxiv.org/abs/1606.03808} {arXiv:1606.03808
  [hep-ph]} \BibitemShut {NoStop}%
\bibitem [{\citenamefont {Chernodub}\ and\ \citenamefont
  {Gongyo}(2017{\natexlab{a}})}]{Chernodub:2016kxh}%
  \BibitemOpen
  \bibfield  {author} {\bibinfo {author} {\bibfnamefont {M.~N.}\ \bibnamefont
  {Chernodub}}\ and\ \bibinfo {author} {\bibfnamefont {S.}~\bibnamefont
  {Gongyo}},\ }\bibfield  {title} {\bibinfo {title} {{Interacting fermions in
  rotation: chiral symmetry restoration, moment of inertia and
  thermodynamics}},\ }\href {https://doi.org/10.1007/JHEP01(2017)136}
  {\bibfield  {journal} {\bibinfo  {journal} {JHEP}\ }\textbf {\bibinfo
  {volume} {01}},\ \bibinfo {pages} {136}},\ \Eprint
  {https://arxiv.org/abs/1611.02598} {arXiv:1611.02598 [hep-th]} \BibitemShut
  {NoStop}%
\bibitem [{\citenamefont {Vilenkin}(1980)}]{Vilenkin:1980zv}%
  \BibitemOpen
  \bibfield  {author} {\bibinfo {author} {\bibfnamefont {A.}~\bibnamefont
  {Vilenkin}},\ }\bibfield  {title} {\bibinfo {title} {{QUANTUM FIELD THEORY AT
  FINITE TEMPERATURE IN A ROTATING SYSTEM}},\ }\href
  {https://doi.org/10.1103/PhysRevD.21.2260} {\bibfield  {journal} {\bibinfo
  {journal} {Phys. Rev. D}\ }\textbf {\bibinfo {volume} {21}},\ \bibinfo
  {pages} {2260} (\bibinfo {year} {1980})}\BibitemShut {NoStop}%
\bibitem [{\citenamefont {Ebihara}\ \emph {et~al.}(2017)\citenamefont
  {Ebihara}, \citenamefont {Fukushima},\ and\ \citenamefont
  {Mameda}}]{Ebihara:2016fwa}%
  \BibitemOpen
  \bibfield  {author} {\bibinfo {author} {\bibfnamefont {S.}~\bibnamefont
  {Ebihara}}, \bibinfo {author} {\bibfnamefont {K.}~\bibnamefont {Fukushima}},\
  and\ \bibinfo {author} {\bibfnamefont {K.}~\bibnamefont {Mameda}},\
  }\bibfield  {title} {\bibinfo {title} {{Boundary effects and gapped
  dispersion in rotating fermionic matter}},\ }\href
  {https://doi.org/10.1016/j.physletb.2016.11.010} {\bibfield  {journal}
  {\bibinfo  {journal} {Phys. Lett. B}\ }\textbf {\bibinfo {volume} {764}},\
  \bibinfo {pages} {94} (\bibinfo {year} {2017})},\ \Eprint
  {https://arxiv.org/abs/1608.00336} {arXiv:1608.00336 [hep-ph]} \BibitemShut
  {NoStop}%
\bibitem [{\citenamefont {Chen}\ \emph {et~al.}(2017)\citenamefont {Chen},
  \citenamefont {Fukushima}, \citenamefont {Huang},\ and\ \citenamefont
  {Mameda}}]{Chen:2017xrj}%
  \BibitemOpen
  \bibfield  {author} {\bibinfo {author} {\bibfnamefont {H.-L.}\ \bibnamefont
  {Chen}}, \bibinfo {author} {\bibfnamefont {K.}~\bibnamefont {Fukushima}},
  \bibinfo {author} {\bibfnamefont {X.-G.}\ \bibnamefont {Huang}},\ and\
  \bibinfo {author} {\bibfnamefont {K.}~\bibnamefont {Mameda}},\ }\bibfield
  {title} {\bibinfo {title} {{Surface Magnetic Catalysis}},\ }\href
  {https://doi.org/10.1103/PhysRevD.96.054032} {\bibfield  {journal} {\bibinfo
  {journal} {Phys. Rev. D}\ }\textbf {\bibinfo {volume} {96}},\ \bibinfo
  {pages} {054032} (\bibinfo {year} {2017})},\ \Eprint
  {https://arxiv.org/abs/1707.09130} {arXiv:1707.09130 [hep-ph]} \BibitemShut
  {NoStop}%
\bibitem [{\citenamefont {Chernodub}\ and\ \citenamefont
  {Gongyo}(2017{\natexlab{b}})}]{Chernodub:2017ref}%
  \BibitemOpen
  \bibfield  {author} {\bibinfo {author} {\bibfnamefont {M.~N.}\ \bibnamefont
  {Chernodub}}\ and\ \bibinfo {author} {\bibfnamefont {S.}~\bibnamefont
  {Gongyo}},\ }\bibfield  {title} {\bibinfo {title} {{Effects of rotation and
  boundaries on chiral symmetry breaking of relativistic fermions}},\ }\href
  {https://doi.org/10.1103/PhysRevD.95.096006} {\bibfield  {journal} {\bibinfo
  {journal} {Phys. Rev. D}\ }\textbf {\bibinfo {volume} {95}},\ \bibinfo
  {pages} {096006} (\bibinfo {year} {2017}{\natexlab{b}})},\ \Eprint
  {https://arxiv.org/abs/1702.08266} {arXiv:1702.08266 [hep-th]} \BibitemShut
  {NoStop}%
\bibitem [{\citenamefont {Wang}\ \emph
  {et~al.}(2019{\natexlab{a}})\citenamefont {Wang}, \citenamefont {Wei},
  \citenamefont {Li},\ and\ \citenamefont {Huang}}]{Wang:2018sur}%
  \BibitemOpen
  \bibfield  {author} {\bibinfo {author} {\bibfnamefont {X.}~\bibnamefont
  {Wang}}, \bibinfo {author} {\bibfnamefont {M.}~\bibnamefont {Wei}}, \bibinfo
  {author} {\bibfnamefont {Z.}~\bibnamefont {Li}},\ and\ \bibinfo {author}
  {\bibfnamefont {M.}~\bibnamefont {Huang}},\ }\bibfield  {title} {\bibinfo
  {title} {{Quark matter under rotation in the NJL model with vector
  interaction}},\ }\href {https://doi.org/10.1103/PhysRevD.99.016018}
  {\bibfield  {journal} {\bibinfo  {journal} {Phys. Rev. D}\ }\textbf {\bibinfo
  {volume} {99}},\ \bibinfo {pages} {016018} (\bibinfo {year}
  {2019}{\natexlab{a}})},\ \Eprint {https://arxiv.org/abs/1808.01931}
  {arXiv:1808.01931 [hep-ph]} \BibitemShut {NoStop}%
\bibitem [{\citenamefont {Wang}\ \emph
  {et~al.}(2019{\natexlab{b}})\citenamefont {Wang}, \citenamefont {Jiang},
  \citenamefont {He},\ and\ \citenamefont {Zhuang}}]{Wang:2018zrn}%
  \BibitemOpen
  \bibfield  {author} {\bibinfo {author} {\bibfnamefont {L.}~\bibnamefont
  {Wang}}, \bibinfo {author} {\bibfnamefont {Y.}~\bibnamefont {Jiang}},
  \bibinfo {author} {\bibfnamefont {L.}~\bibnamefont {He}},\ and\ \bibinfo
  {author} {\bibfnamefont {P.}~\bibnamefont {Zhuang}},\ }\bibfield  {title}
  {\bibinfo {title} {{Local suppression and enhancement of the pairing
  condensate under rotation}},\ }\href
  {https://doi.org/10.1103/PhysRevC.100.034902} {\bibfield  {journal} {\bibinfo
   {journal} {Phys. Rev. C}\ }\textbf {\bibinfo {volume} {100}},\ \bibinfo
  {pages} {034902} (\bibinfo {year} {2019}{\natexlab{b}})},\ \Eprint
  {https://arxiv.org/abs/1901.00804} {arXiv:1901.00804 [nucl-th]} \BibitemShut
  {NoStop}%
\bibitem [{\citenamefont {Wang}\ \emph
  {et~al.}(2019{\natexlab{c}})\citenamefont {Wang}, \citenamefont {Jiang},
  \citenamefont {He},\ and\ \citenamefont {Zhuang}}]{Wang:2019nhd}%
  \BibitemOpen
  \bibfield  {author} {\bibinfo {author} {\bibfnamefont {L.}~\bibnamefont
  {Wang}}, \bibinfo {author} {\bibfnamefont {Y.}~\bibnamefont {Jiang}},
  \bibinfo {author} {\bibfnamefont {L.}~\bibnamefont {He}},\ and\ \bibinfo
  {author} {\bibfnamefont {P.}~\bibnamefont {Zhuang}},\ }\bibfield  {title}
  {\bibinfo {title} {{Chiral vortices and pseudoscalar condensation due to
  rotation}},\ }\href {https://doi.org/10.1103/PhysRevD.100.114009} {\bibfield
  {journal} {\bibinfo  {journal} {Phys. Rev. D}\ }\textbf {\bibinfo {volume}
  {100}},\ \bibinfo {pages} {114009} (\bibinfo {year} {2019}{\natexlab{c}})},\
  \Eprint {https://arxiv.org/abs/1901.04697} {arXiv:1901.04697 [nucl-th]}
  \BibitemShut {NoStop}%
\bibitem [{\citenamefont {Zhang}\ \emph
  {et~al.}(2020{\natexlab{a}})\citenamefont {Zhang}, \citenamefont {Shi},
  \citenamefont {Luo},\ and\ \citenamefont {Zong}}]{Zhang:2020jux}%
  \BibitemOpen
  \bibfield  {author} {\bibinfo {author} {\bibfnamefont {Z.}~\bibnamefont
  {Zhang}}, \bibinfo {author} {\bibfnamefont {C.}~\bibnamefont {Shi}}, \bibinfo
  {author} {\bibfnamefont {X.}~\bibnamefont {Luo}},\ and\ \bibinfo {author}
  {\bibfnamefont {H.-S.}\ \bibnamefont {Zong}},\ }\bibfield  {title} {\bibinfo
  {title} {{Chiral phase transition in a rotating sphere}},\ }\href
  {https://doi.org/10.1103/PhysRevD.101.074036} {\bibfield  {journal} {\bibinfo
   {journal} {Phys. Rev. D}\ }\textbf {\bibinfo {volume} {101}},\ \bibinfo
  {pages} {074036} (\bibinfo {year} {2020}{\natexlab{a}})},\ \Eprint
  {https://arxiv.org/abs/2003.03765} {arXiv:2003.03765 [nucl-th]} \BibitemShut
  {NoStop}%
\bibitem [{\citenamefont {Sadooghi}\ \emph {et~al.}(2021)\citenamefont
  {Sadooghi}, \citenamefont {Tabatabaee~Mehr},\ and\ \citenamefont
  {Taghinavaz}}]{Sadooghi:2021upd}%
  \BibitemOpen
  \bibfield  {author} {\bibinfo {author} {\bibfnamefont {N.}~\bibnamefont
  {Sadooghi}}, \bibinfo {author} {\bibfnamefont {S.~M.~A.}\ \bibnamefont
  {Tabatabaee~Mehr}},\ and\ \bibinfo {author} {\bibfnamefont {F.}~\bibnamefont
  {Taghinavaz}},\ }\bibfield  {title} {\bibinfo {title} {{Inverse
  magnetorotational catalysis and the phase diagram of a rotating hot and
  magnetized quark matter}},\ }\href
  {https://doi.org/10.1103/PhysRevD.104.116022} {\bibfield  {journal} {\bibinfo
   {journal} {Phys. Rev. D}\ }\textbf {\bibinfo {volume} {104}},\ \bibinfo
  {pages} {116022} (\bibinfo {year} {2021})},\ \Eprint
  {https://arxiv.org/abs/2108.12760} {arXiv:2108.12760 [hep-ph]} \BibitemShut
  {NoStop}%
\bibitem [{\citenamefont {Chen}\ \emph
  {et~al.}(2022{\natexlab{a}})\citenamefont {Chen}, \citenamefont {Li},\ and\
  \citenamefont {Huang}}]{Chen:2022mhf}%
  \BibitemOpen
  \bibfield  {author} {\bibinfo {author} {\bibfnamefont {Y.}~\bibnamefont
  {Chen}}, \bibinfo {author} {\bibfnamefont {D.}~\bibnamefont {Li}},\ and\
  \bibinfo {author} {\bibfnamefont {M.}~\bibnamefont {Huang}},\ }\bibfield
  {title} {\bibinfo {title} {{Inhomogeneous chiral condensation under rotation
  in the holographic QCD}},\ }\href
  {https://doi.org/10.1103/PhysRevD.106.106002} {\bibfield  {journal} {\bibinfo
   {journal} {Phys. Rev. D}\ }\textbf {\bibinfo {volume} {106}},\ \bibinfo
  {pages} {106002} (\bibinfo {year} {2022}{\natexlab{a}})},\ \Eprint
  {https://arxiv.org/abs/2208.05668} {arXiv:2208.05668 [hep-ph]} \BibitemShut
  {NoStop}%
\bibitem [{\citenamefont {Chen}\ \emph {et~al.}(2023)\citenamefont {Chen},
  \citenamefont {Zhu},\ and\ \citenamefont {Huang}}]{Chen:2023cjt}%
  \BibitemOpen
  \bibfield  {author} {\bibinfo {author} {\bibfnamefont {H.-L.}\ \bibnamefont
  {Chen}}, \bibinfo {author} {\bibfnamefont {Z.-B.}\ \bibnamefont {Zhu}},\ and\
  \bibinfo {author} {\bibfnamefont {X.-G.}\ \bibnamefont {Huang}},\ }\bibfield
  {title} {\bibinfo {title} {{Quark-meson model under rotation: A functional
  renormalization group study}},\ }\href@noop {} {\  (\bibinfo {year}
  {2023})},\ \Eprint {https://arxiv.org/abs/2306.08362} {arXiv:2306.08362
  [hep-ph]} \BibitemShut {NoStop}%
\bibitem [{\citenamefont {Huang}\ \emph {et~al.}(2018)\citenamefont {Huang},
  \citenamefont {Nishimura},\ and\ \citenamefont {Yamamoto}}]{Huang:2017pqe}%
  \BibitemOpen
  \bibfield  {author} {\bibinfo {author} {\bibfnamefont {X.-G.}\ \bibnamefont
  {Huang}}, \bibinfo {author} {\bibfnamefont {K.}~\bibnamefont {Nishimura}},\
  and\ \bibinfo {author} {\bibfnamefont {N.}~\bibnamefont {Yamamoto}},\
  }\bibfield  {title} {\bibinfo {title} {{Anomalous effects of dense matter
  under rotation}},\ }\href {https://doi.org/10.1007/JHEP02(2018)069}
  {\bibfield  {journal} {\bibinfo  {journal} {JHEP}\ }\textbf {\bibinfo
  {volume} {02}},\ \bibinfo {pages} {069}},\ \Eprint
  {https://arxiv.org/abs/1711.02190} {arXiv:1711.02190 [hep-ph]} \BibitemShut
  {NoStop}%
\bibitem [{\citenamefont {Liu}\ and\ \citenamefont
  {Zahed}(2018)}]{Liu:2017spl}%
  \BibitemOpen
  \bibfield  {author} {\bibinfo {author} {\bibfnamefont {Y.}~\bibnamefont
  {Liu}}\ and\ \bibinfo {author} {\bibfnamefont {I.}~\bibnamefont {Zahed}},\
  }\bibfield  {title} {\bibinfo {title} {{Pion Condensation by Rotation in a
  Magnetic field}},\ }\href {https://doi.org/10.1103/PhysRevLett.120.032001}
  {\bibfield  {journal} {\bibinfo  {journal} {Phys. Rev. Lett.}\ }\textbf
  {\bibinfo {volume} {120}},\ \bibinfo {pages} {032001} (\bibinfo {year}
  {2018})},\ \Eprint {https://arxiv.org/abs/1711.08354} {arXiv:1711.08354
  [hep-ph]} \BibitemShut {NoStop}%
\bibitem [{\citenamefont {Zhang}\ \emph
  {et~al.}(2020{\natexlab{b}})\citenamefont {Zhang}, \citenamefont {Hou},\ and\
  \citenamefont {Liao}}]{Zhang:2018ome}%
  \BibitemOpen
  \bibfield  {author} {\bibinfo {author} {\bibfnamefont {H.}~\bibnamefont
  {Zhang}}, \bibinfo {author} {\bibfnamefont {D.}~\bibnamefont {Hou}},\ and\
  \bibinfo {author} {\bibfnamefont {J.}~\bibnamefont {Liao}},\ }\bibfield
  {title} {\bibinfo {title} {{Mesonic Condensation in Isospin Matter under
  Rotation}},\ }\href {https://doi.org/10.1088/1674-1137/abae4d} {\bibfield
  {journal} {\bibinfo  {journal} {Chin. Phys. C}\ }\textbf {\bibinfo {volume}
  {44}},\ \bibinfo {pages} {111001} (\bibinfo {year} {2020}{\natexlab{b}})},\
  \Eprint {https://arxiv.org/abs/1812.11787} {arXiv:1812.11787 [hep-ph]}
  \BibitemShut {NoStop}%
\bibitem [{\citenamefont {Chen}\ \emph {et~al.}(2019)\citenamefont {Chen},
  \citenamefont {Huang},\ and\ \citenamefont {Mameda}}]{Chen:2019tcp}%
  \BibitemOpen
  \bibfield  {author} {\bibinfo {author} {\bibfnamefont {H.-L.}\ \bibnamefont
  {Chen}}, \bibinfo {author} {\bibfnamefont {X.-G.}\ \bibnamefont {Huang}},\
  and\ \bibinfo {author} {\bibfnamefont {K.}~\bibnamefont {Mameda}},\
  }\bibfield  {title} {\bibinfo {title} {{Do charged pions condense in a
  magnetic field with rotation?}},\ }\href@noop {} {\  (\bibinfo {year}
  {2019})},\ \Eprint {https://arxiv.org/abs/1910.02700} {arXiv:1910.02700
  [nucl-th]} \BibitemShut {NoStop}%
\bibitem [{\citenamefont {Cao}\ and\ \citenamefont {He}(2019)}]{Cao:2019ctl}%
  \BibitemOpen
  \bibfield  {author} {\bibinfo {author} {\bibfnamefont {G.}~\bibnamefont
  {Cao}}\ and\ \bibinfo {author} {\bibfnamefont {L.}~\bibnamefont {He}},\
  }\bibfield  {title} {\bibinfo {title} {{Rotation induced charged pion
  condensation in a strong magnetic field: A Nambu\textendash{}Jona-Lasino
  model study}},\ }\href {https://doi.org/10.1103/PhysRevD.100.094015}
  {\bibfield  {journal} {\bibinfo  {journal} {Phys. Rev. D}\ }\textbf {\bibinfo
  {volume} {100}},\ \bibinfo {pages} {094015} (\bibinfo {year} {2019})},\
  \Eprint {https://arxiv.org/abs/1910.02728} {arXiv:1910.02728 [nucl-th]}
  \BibitemShut {NoStop}%
\bibitem [{\citenamefont {Nishimura}\ and\ \citenamefont
  {Yamamoto}(2020)}]{Nishimura:2020odq}%
  \BibitemOpen
  \bibfield  {author} {\bibinfo {author} {\bibfnamefont {K.}~\bibnamefont
  {Nishimura}}\ and\ \bibinfo {author} {\bibfnamefont {N.}~\bibnamefont
  {Yamamoto}},\ }\bibfield  {title} {\bibinfo {title} {{Topological term, QCD
  anomaly, and the $\eta^{'}$ chiral soliton lattice in rotating baryonic
  matter}},\ }\href {https://doi.org/10.1007/JHEP07(2020)196} {\bibfield
  {journal} {\bibinfo  {journal} {JHEP}\ }\textbf {\bibinfo {volume}
  {07}}\bibfield  {number} {\bibinfo  {number} { (07)},\ \bibinfo {pages}
  {196}},\ }\Eprint {https://arxiv.org/abs/2003.13945} {arXiv:2003.13945
  [hep-ph]} \BibitemShut {NoStop}%
\bibitem [{\citenamefont {Chen}\ \emph
  {et~al.}(2021{\natexlab{b}})\citenamefont {Chen}, \citenamefont {Zhang},
  \citenamefont {Li}, \citenamefont {Hou},\ and\ \citenamefont
  {Huang}}]{Chen:2020ath}%
  \BibitemOpen
  \bibfield  {author} {\bibinfo {author} {\bibfnamefont {X.}~\bibnamefont
  {Chen}}, \bibinfo {author} {\bibfnamefont {L.}~\bibnamefont {Zhang}},
  \bibinfo {author} {\bibfnamefont {D.}~\bibnamefont {Li}}, \bibinfo {author}
  {\bibfnamefont {D.}~\bibnamefont {Hou}},\ and\ \bibinfo {author}
  {\bibfnamefont {M.}~\bibnamefont {Huang}},\ }\bibfield  {title} {\bibinfo
  {title} {{Gluodynamics and deconfinement phase transition under rotation from
  holography}},\ }\href {https://doi.org/10.1007/JHEP07(2021)132} {\bibfield
  {journal} {\bibinfo  {journal} {JHEP}\ }\textbf {\bibinfo {volume} {07}},\
  \bibinfo {pages} {132}},\ \Eprint {https://arxiv.org/abs/2010.14478}
  {arXiv:2010.14478 [hep-ph]} \BibitemShut {NoStop}%
\bibitem [{\citenamefont {Braga}\ \emph {et~al.}(2022)\citenamefont {Braga},
  \citenamefont {Faulhaber},\ and\ \citenamefont {Junqueira}}]{Braga:2022yfe}%
  \BibitemOpen
  \bibfield  {author} {\bibinfo {author} {\bibfnamefont {N.~R.~F.}\
  \bibnamefont {Braga}}, \bibinfo {author} {\bibfnamefont {L.~F.}\ \bibnamefont
  {Faulhaber}},\ and\ \bibinfo {author} {\bibfnamefont {O.~C.}\ \bibnamefont
  {Junqueira}},\ }\bibfield  {title} {\bibinfo {title}
  {{Confinement-deconfinement temperature for a rotating quark-gluon plasma}},\
  }\href {https://doi.org/10.1103/PhysRevD.105.106003} {\bibfield  {journal}
  {\bibinfo  {journal} {Phys. Rev. D}\ }\textbf {\bibinfo {volume} {105}},\
  \bibinfo {pages} {106003} (\bibinfo {year} {2022})},\ \Eprint
  {https://arxiv.org/abs/2201.05581} {arXiv:2201.05581 [hep-th]} \BibitemShut
  {NoStop}%
\bibitem [{\citenamefont {Yadav}(2022)}]{Yadav:2022qcl}%
  \BibitemOpen
  \bibfield  {author} {\bibinfo {author} {\bibfnamefont {G.}~\bibnamefont
  {Yadav}},\ }\bibfield  {title} {\bibinfo {title} {{Deconfinement Temperature
  of Rotating QGP at Intermediate Coupling from ${\cal M}$-Theory}},\
  }\href@noop {} {\  (\bibinfo {year} {2022})},\ \Eprint
  {https://arxiv.org/abs/2203.11959} {arXiv:2203.11959 [hep-th]} \BibitemShut
  {NoStop}%
\bibitem [{\citenamefont {Zhao}\ \emph {et~al.}(2022)\citenamefont {Zhao},
  \citenamefont {He}, \citenamefont {Hou}, \citenamefont {Li},\ and\
  \citenamefont {Li}}]{Zhao:2022uxc}%
  \BibitemOpen
  \bibfield  {author} {\bibinfo {author} {\bibfnamefont {Y.-Q.}\ \bibnamefont
  {Zhao}}, \bibinfo {author} {\bibfnamefont {S.}~\bibnamefont {He}}, \bibinfo
  {author} {\bibfnamefont {D.}~\bibnamefont {Hou}}, \bibinfo {author}
  {\bibfnamefont {L.}~\bibnamefont {Li}},\ and\ \bibinfo {author}
  {\bibfnamefont {Z.}~\bibnamefont {Li}},\ }\bibfield  {title} {\bibinfo
  {title} {{Phase diagram of holographic thermal dense QCD matter with
  rotation}},\ }\href@noop {} {\  (\bibinfo {year} {2022})},\ \Eprint
  {https://arxiv.org/abs/2212.14662} {arXiv:2212.14662 [hep-ph]} \BibitemShut
  {NoStop}%
\bibitem [{\citenamefont {Fujimoto}\ \emph {et~al.}(2021)\citenamefont
  {Fujimoto}, \citenamefont {Fukushima},\ and\ \citenamefont
  {Hidaka}}]{Fujimoto:2021xix}%
  \BibitemOpen
  \bibfield  {author} {\bibinfo {author} {\bibfnamefont {Y.}~\bibnamefont
  {Fujimoto}}, \bibinfo {author} {\bibfnamefont {K.}~\bibnamefont
  {Fukushima}},\ and\ \bibinfo {author} {\bibfnamefont {Y.}~\bibnamefont
  {Hidaka}},\ }\bibfield  {title} {\bibinfo {title} {{Deconfining Phase
  Boundary of Rapidly Rotating Hot and Dense Matter and Analysis of Moment of
  Inertia}},\ }\href {https://doi.org/10.1016/j.physletb.2021.136184}
  {\bibfield  {journal} {\bibinfo  {journal} {Phys. Lett. B}\ }\textbf
  {\bibinfo {volume} {816}},\ \bibinfo {pages} {136184} (\bibinfo {year}
  {2021})},\ \Eprint {https://arxiv.org/abs/2101.09173} {arXiv:2101.09173
  [hep-ph]} \BibitemShut {NoStop}%
\bibitem [{\citenamefont {Chen}\ \emph
  {et~al.}(2022{\natexlab{b}})\citenamefont {Chen}, \citenamefont {Fukushima},\
  and\ \citenamefont {Shimada}}]{Chen:2022smf}%
  \BibitemOpen
  \bibfield  {author} {\bibinfo {author} {\bibfnamefont {S.}~\bibnamefont
  {Chen}}, \bibinfo {author} {\bibfnamefont {K.}~\bibnamefont {Fukushima}},\
  and\ \bibinfo {author} {\bibfnamefont {Y.}~\bibnamefont {Shimada}},\
  }\bibfield  {title} {\bibinfo {title} {{Perturbative Confinement in Thermal
  Yang-Mills Theories Induced by Imaginary Angular Velocity}},\ }\href
  {https://doi.org/10.1103/PhysRevLett.129.242002} {\bibfield  {journal}
  {\bibinfo  {journal} {Phys. Rev. Lett.}\ }\textbf {\bibinfo {volume} {129}},\
  \bibinfo {pages} {242002} (\bibinfo {year} {2022}{\natexlab{b}})},\ \Eprint
  {https://arxiv.org/abs/2207.12665} {arXiv:2207.12665 [hep-ph]} \BibitemShut
  {NoStop}%
\bibitem [{\citenamefont {Chernodub}\ \emph {et~al.}(2022)\citenamefont
  {Chernodub}, \citenamefont {Goy},\ and\ \citenamefont
  {Molochkov}}]{Chernodub:2022veq}%
  \BibitemOpen
  \bibfield  {author} {\bibinfo {author} {\bibfnamefont {M.~N.}\ \bibnamefont
  {Chernodub}}, \bibinfo {author} {\bibfnamefont {V.~A.}\ \bibnamefont {Goy}},\
  and\ \bibinfo {author} {\bibfnamefont {A.~V.}\ \bibnamefont {Molochkov}},\
  }\bibfield  {title} {\bibinfo {title} {{Inhomogeneity of rotating gluon
  plasma and Tolman-Ehrenfest law in imaginary time: lattice results for fast
  imaginary rotation}},\ }\href@noop {} {\  (\bibinfo {year} {2022})},\ \Eprint
  {https://arxiv.org/abs/2209.15534} {arXiv:2209.15534 [hep-lat]} \BibitemShut
  {NoStop}%
\bibitem [{\citenamefont {Braguta}\ \emph {et~al.}(2020)\citenamefont
  {Braguta}, \citenamefont {Kotov}, \citenamefont {Kuznedelev},\ and\
  \citenamefont {Roenko}}]{Braguta:2020biu}%
  \BibitemOpen
  \bibfield  {author} {\bibinfo {author} {\bibfnamefont {V.~V.}\ \bibnamefont
  {Braguta}}, \bibinfo {author} {\bibfnamefont {A.~Y.}\ \bibnamefont {Kotov}},
  \bibinfo {author} {\bibfnamefont {D.~D.}\ \bibnamefont {Kuznedelev}},\ and\
  \bibinfo {author} {\bibfnamefont {A.~A.}\ \bibnamefont {Roenko}},\ }\bibfield
   {title} {\bibinfo {title} {{Study of the Confinement/Deconfinement Phase
  Transition in Rotating Lattice SU(3) Gluodynamics}},\ }\href
  {https://doi.org/10.31857/S1234567820130029} {\bibfield  {journal} {\bibinfo
  {journal} {Pisma Zh. Eksp. Teor. Fiz.}\ }\textbf {\bibinfo {volume} {112}},\
  \bibinfo {pages} {9} (\bibinfo {year} {2020})}\BibitemShut {NoStop}%
\bibitem [{\citenamefont {Braguta}\ \emph {et~al.}(2021)\citenamefont
  {Braguta}, \citenamefont {Kotov}, \citenamefont {Kuznedelev},\ and\
  \citenamefont {Roenko}}]{Braguta:2021jgn}%
  \BibitemOpen
  \bibfield  {author} {\bibinfo {author} {\bibfnamefont {V.~V.}\ \bibnamefont
  {Braguta}}, \bibinfo {author} {\bibfnamefont {A.~Y.}\ \bibnamefont {Kotov}},
  \bibinfo {author} {\bibfnamefont {D.~D.}\ \bibnamefont {Kuznedelev}},\ and\
  \bibinfo {author} {\bibfnamefont {A.~A.}\ \bibnamefont {Roenko}},\ }\bibfield
   {title} {\bibinfo {title} {{Influence of relativistic rotation on the
  confinement-deconfinement transition in gluodynamics}},\ }\href
  {https://doi.org/10.1103/PhysRevD.103.094515} {\bibfield  {journal} {\bibinfo
   {journal} {Phys. Rev. D}\ }\textbf {\bibinfo {volume} {103}},\ \bibinfo
  {pages} {094515} (\bibinfo {year} {2021})},\ \Eprint
  {https://arxiv.org/abs/2102.05084} {arXiv:2102.05084 [hep-lat]} \BibitemShut
  {NoStop}%
\bibitem [{\citenamefont {Yamamoto}\ and\ \citenamefont
  {Hirono}(2013)}]{Yamamoto:2013zwa}%
  \BibitemOpen
  \bibfield  {author} {\bibinfo {author} {\bibfnamefont {A.}~\bibnamefont
  {Yamamoto}}\ and\ \bibinfo {author} {\bibfnamefont {Y.}~\bibnamefont
  {Hirono}},\ }\bibfield  {title} {\bibinfo {title} {{Lattice QCD in rotating
  frames}},\ }\href {https://doi.org/10.1103/PhysRevLett.111.081601} {\bibfield
   {journal} {\bibinfo  {journal} {Phys. Rev. Lett.}\ }\textbf {\bibinfo
  {volume} {111}},\ \bibinfo {pages} {081601} (\bibinfo {year} {2013})},\
  \Eprint {https://arxiv.org/abs/1303.6292} {arXiv:1303.6292 [hep-lat]}
  \BibitemShut {NoStop}%
\bibitem [{\citenamefont {Rothe}(2012)}]{Rothe:1992nt}%
  \BibitemOpen
  \bibfield  {author} {\bibinfo {author} {\bibfnamefont {H.~J.}\ \bibnamefont
  {Rothe}},\ }\href {https://doi.org/10.1142/8229} {\emph {\bibinfo {title}
  {{Lattice Gauge Theories : An Introduction (Fourth Edition)}}}},\
  Vol.~\bibinfo {volume} {43}\ (\bibinfo  {publisher} {World Scientific
  Publishing Company},\ \bibinfo {year} {2012})\BibitemShut {NoStop}%
\bibitem [{\citenamefont {Tiburzi}(2013)}]{Tiburzi:2013vza}%
  \BibitemOpen
  \bibfield  {author} {\bibinfo {author} {\bibfnamefont {B.~C.}\ \bibnamefont
  {Tiburzi}},\ }\bibfield  {title} {\bibinfo {title} {{Chiral Symmetry
  Restoration from a Boundary}},\ }\href
  {https://doi.org/10.1103/PhysRevD.88.034027} {\bibfield  {journal} {\bibinfo
  {journal} {Phys. Rev. D}\ }\textbf {\bibinfo {volume} {88}},\ \bibinfo
  {pages} {034027} (\bibinfo {year} {2013})},\ \Eprint
  {https://arxiv.org/abs/1302.6645} {arXiv:1302.6645 [hep-lat]} \BibitemShut
  {NoStop}%
\bibitem [{\citenamefont {Ji}\ \emph {et~al.}(2021)\citenamefont {Ji},
  \citenamefont {Yuan},\ and\ \citenamefont {Zhao}}]{Ji:2020ena}%
  \BibitemOpen
  \bibfield  {author} {\bibinfo {author} {\bibfnamefont {X.}~\bibnamefont
  {Ji}}, \bibinfo {author} {\bibfnamefont {F.}~\bibnamefont {Yuan}},\ and\
  \bibinfo {author} {\bibfnamefont {Y.}~\bibnamefont {Zhao}},\ }\bibfield
  {title} {\bibinfo {title} {{What we know and what we don\textquoteright{}t
  know about the proton spin after 30 years}},\ }\href
  {https://doi.org/10.1038/s42254-020-00248-4} {\bibfield  {journal} {\bibinfo
  {journal} {Nature Rev. Phys.}\ }\textbf {\bibinfo {volume} {3}},\ \bibinfo
  {pages} {27} (\bibinfo {year} {2021})},\ \Eprint
  {https://arxiv.org/abs/2009.01291} {arXiv:2009.01291 [hep-ph]} \BibitemShut
  {NoStop}%
\bibitem [{\citenamefont {Ji}(1997)}]{Ji:1996ek}%
  \BibitemOpen
  \bibfield  {author} {\bibinfo {author} {\bibfnamefont {X.-D.}\ \bibnamefont
  {Ji}},\ }\bibfield  {title} {\bibinfo {title} {{Gauge-Invariant Decomposition
  of Nucleon Spin}},\ }\href {https://doi.org/10.1103/PhysRevLett.78.610}
  {\bibfield  {journal} {\bibinfo  {journal} {Phys. Rev. Lett.}\ }\textbf
  {\bibinfo {volume} {78}},\ \bibinfo {pages} {610} (\bibinfo {year} {1997})},\
  \Eprint {https://arxiv.org/abs/hep-ph/9603249} {arXiv:hep-ph/9603249}
  \BibitemShut {NoStop}%
\bibitem [{\citenamefont {Braguta}\ \emph
  {et~al.}(2023{\natexlab{a}})\citenamefont {Braguta}, \citenamefont
  {Chernodub}, \citenamefont {Roenko},\ and\ \citenamefont
  {Sychev}}]{Braguta:2023yjn}%
  \BibitemOpen
  \bibfield  {author} {\bibinfo {author} {\bibfnamefont {V.~V.}\ \bibnamefont
  {Braguta}}, \bibinfo {author} {\bibfnamefont {M.~N.}\ \bibnamefont
  {Chernodub}}, \bibinfo {author} {\bibfnamefont {A.~A.}\ \bibnamefont
  {Roenko}},\ and\ \bibinfo {author} {\bibfnamefont {D.~A.}\ \bibnamefont
  {Sychev}},\ }\bibfield  {title} {\bibinfo {title} {{Negative moment of
  inertia and rotational instability of gluon plasma}},\ }\href@noop {} {\
  (\bibinfo {year} {2023}{\natexlab{a}})},\ \Eprint
  {https://arxiv.org/abs/2303.03147} {arXiv:2303.03147 [hep-lat]} \BibitemShut
  {NoStop}%
\bibitem [{\citenamefont {Cheng}\ \emph {et~al.}(2008)\citenamefont {Cheng}
  \emph {et~al.}}]{Cheng:2007jq}%
  \BibitemOpen
  \bibfield  {author} {\bibinfo {author} {\bibfnamefont {M.}~\bibnamefont
  {Cheng}} \emph {et~al.},\ }\bibfield  {title} {\bibinfo {title} {{The QCD
  equation of state with almost physical quark masses}},\ }\href
  {https://doi.org/10.1103/PhysRevD.77.014511} {\bibfield  {journal} {\bibinfo
  {journal} {Phys. Rev. D}\ }\textbf {\bibinfo {volume} {77}},\ \bibinfo
  {pages} {014511} (\bibinfo {year} {2008})},\ \Eprint
  {https://arxiv.org/abs/0710.0354} {arXiv:0710.0354 [hep-lat]} \BibitemShut
  {NoStop}%
\bibitem [{\citenamefont {Bazavov}\ \emph {et~al.}(2012)\citenamefont {Bazavov}
  \emph {et~al.}}]{Bazavov:2011nk}%
  \BibitemOpen
  \bibfield  {author} {\bibinfo {author} {\bibfnamefont {A.}~\bibnamefont
  {Bazavov}} \emph {et~al.},\ }\bibfield  {title} {\bibinfo {title} {{The
  chiral and deconfinement aspects of the QCD transition}},\ }\href
  {https://doi.org/10.1103/PhysRevD.85.054503} {\bibfield  {journal} {\bibinfo
  {journal} {Phys. Rev. D}\ }\textbf {\bibinfo {volume} {85}},\ \bibinfo
  {pages} {054503} (\bibinfo {year} {2012})},\ \Eprint
  {https://arxiv.org/abs/1111.1710} {arXiv:1111.1710 [hep-lat]} \BibitemShut
  {NoStop}%
\bibitem [{\citenamefont {Cheng}\ \emph {et~al.}(2010)\citenamefont {Cheng}
  \emph {et~al.}}]{Cheng:2009zi}%
  \BibitemOpen
  \bibfield  {author} {\bibinfo {author} {\bibfnamefont {M.}~\bibnamefont
  {Cheng}} \emph {et~al.},\ }\bibfield  {title} {\bibinfo {title} {{Equation of
  State for physical quark masses}},\ }\href
  {https://doi.org/10.1103/PhysRevD.81.054504} {\bibfield  {journal} {\bibinfo
  {journal} {Phys. Rev. D}\ }\textbf {\bibinfo {volume} {81}},\ \bibinfo
  {pages} {054504} (\bibinfo {year} {2010})},\ \Eprint
  {https://arxiv.org/abs/0911.2215} {arXiv:0911.2215 [hep-lat]} \BibitemShut
  {NoStop}%
\bibitem [{\citenamefont {Braguta}\ \emph
  {et~al.}(2023{\natexlab{b}})\citenamefont {Braguta}, \citenamefont {Kotov},
  \citenamefont {Roenko},\ and\ \citenamefont {Sychev}}]{Braguta:2022str}%
  \BibitemOpen
  \bibfield  {author} {\bibinfo {author} {\bibfnamefont {V.~V.}\ \bibnamefont
  {Braguta}}, \bibinfo {author} {\bibfnamefont {A.}~\bibnamefont {Kotov}},
  \bibinfo {author} {\bibfnamefont {A.}~\bibnamefont {Roenko}},\ and\ \bibinfo
  {author} {\bibfnamefont {D.}~\bibnamefont {Sychev}},\ }\bibfield  {title}
  {\bibinfo {title} {{Thermal phase transitions in rotating QCD with dynamical
  quarks}},\ }\href {https://doi.org/10.22323/1.430.0190} {\bibfield  {journal}
  {\bibinfo  {journal} {PoS}\ }\textbf {\bibinfo {volume} {LATTICE2022}},\
  \bibinfo {pages} {190} (\bibinfo {year} {2023}{\natexlab{b}})},\ \Eprint
  {https://arxiv.org/abs/2212.03224} {arXiv:2212.03224 [hep-lat]} \BibitemShut
  {NoStop}%
\bibitem [{\citenamefont {Jiang}(2022)}]{Jiang:2021izj}%
  \BibitemOpen
  \bibfield  {author} {\bibinfo {author} {\bibfnamefont {Y.}~\bibnamefont
  {Jiang}},\ }\bibfield  {title} {\bibinfo {title} {{Chiral vortical
  catalysis}},\ }\href {https://doi.org/10.1140/epjc/s10052-022-10915-8}
  {\bibfield  {journal} {\bibinfo  {journal} {Eur. Phys. J. C}\ }\textbf
  {\bibinfo {volume} {82}},\ \bibinfo {pages} {949} (\bibinfo {year} {2022})},\
  \Eprint {https://arxiv.org/abs/2108.09622} {arXiv:2108.09622 [hep-ph]}
  \BibitemShut {NoStop}%
\bibitem [{\citenamefont {Chernodub}(2022)}]{Chernodub:2022qlz}%
  \BibitemOpen
  \bibfield  {author} {\bibinfo {author} {\bibfnamefont {M.~N.}\ \bibnamefont
  {Chernodub}},\ }\bibfield  {title} {\bibinfo {title} {{Fractal thermodynamics
  and ninionic statistics of coherent rotational states: realization via
  imaginary angular rotation in imaginary time formalism}},\ }\href@noop {} {\
  (\bibinfo {year} {2022})},\ \Eprint {https://arxiv.org/abs/2210.05651}
  {arXiv:2210.05651 [quant-ph]} \BibitemShut {NoStop}%
\bibitem [{\citenamefont {Fradkin}\ and\ \citenamefont
  {Susskind}(1978)}]{Fradkin:1978th}%
  \BibitemOpen
  \bibfield  {author} {\bibinfo {author} {\bibfnamefont {E.~H.}\ \bibnamefont
  {Fradkin}}\ and\ \bibinfo {author} {\bibfnamefont {L.}~\bibnamefont
  {Susskind}},\ }\bibfield  {title} {\bibinfo {title} {{Order and Disorder in
  Gauge Systems and Magnets}},\ }\href
  {https://doi.org/10.1103/PhysRevD.17.2637} {\bibfield  {journal} {\bibinfo
  {journal} {Phys. Rev. D}\ }\textbf {\bibinfo {volume} {17}},\ \bibinfo
  {pages} {2637} (\bibinfo {year} {1978})}\BibitemShut {NoStop}%
\bibitem [{\citenamefont {Kogut}(1979)}]{Kogut:1979wt}%
  \BibitemOpen
  \bibfield  {author} {\bibinfo {author} {\bibfnamefont {J.~B.}\ \bibnamefont
  {Kogut}},\ }\bibfield  {title} {\bibinfo {title} {{An Introduction to Lattice
  Gauge Theory and Spin Systems}},\ }\href
  {https://doi.org/10.1103/RevModPhys.51.659} {\bibfield  {journal} {\bibinfo
  {journal} {Rev. Mod. Phys.}\ }\textbf {\bibinfo {volume} {51}},\ \bibinfo
  {pages} {659} (\bibinfo {year} {1979})}\BibitemShut {NoStop}%
\bibitem [{\citenamefont {Sommer}(1994)}]{Sommer:1993ce}%
  \BibitemOpen
  \bibfield  {author} {\bibinfo {author} {\bibfnamefont {R.}~\bibnamefont
  {Sommer}},\ }\bibfield  {title} {\bibinfo {title} {{A New way to set the
  energy scale in lattice gauge theories and its applications to the static
  force and alpha-s in SU(2) Yang-Mills theory}},\ }\href
  {https://doi.org/10.1016/0550-3213(94)90473-1} {\bibfield  {journal}
  {\bibinfo  {journal} {Nucl. Phys. B}\ }\textbf {\bibinfo {volume} {411}},\
  \bibinfo {pages} {839} (\bibinfo {year} {1994})},\ \Eprint
  {https://arxiv.org/abs/hep-lat/9310022} {arXiv:hep-lat/9310022} \BibitemShut
  {NoStop}%
\bibitem [{\citenamefont {Bali}\ and\ \citenamefont
  {Schilling}(1992)}]{Bali:1992ab}%
  \BibitemOpen
  \bibfield  {author} {\bibinfo {author} {\bibfnamefont {G.~S.}\ \bibnamefont
  {Bali}}\ and\ \bibinfo {author} {\bibfnamefont {K.}~\bibnamefont
  {Schilling}},\ }\bibfield  {title} {\bibinfo {title} {{Static quark -
  anti-quark potential: Scaling behavior and finite size effects in SU(3)
  lattice gauge theory}},\ }\href {https://doi.org/10.1103/PhysRevD.46.2636}
  {\bibfield  {journal} {\bibinfo  {journal} {Phys. Rev. D}\ }\textbf {\bibinfo
  {volume} {46}},\ \bibinfo {pages} {2636} (\bibinfo {year}
  {1992})}\BibitemShut {NoStop}%
\bibitem [{\citenamefont {Bali}\ \emph {et~al.}(2000)\citenamefont {Bali},
  \citenamefont {Bolder}, \citenamefont {Eicker}, \citenamefont {Lippert},
  \citenamefont {Orth}, \citenamefont {Ueberholz}, \citenamefont {Schilling},\
  and\ \citenamefont {Struckmann}}]{Bali:2000vr}%
  \BibitemOpen
  \bibfield  {author} {\bibinfo {author} {\bibfnamefont {G.~S.}\ \bibnamefont
  {Bali}}, \bibinfo {author} {\bibfnamefont {B.}~\bibnamefont {Bolder}},
  \bibinfo {author} {\bibfnamefont {N.}~\bibnamefont {Eicker}}, \bibinfo
  {author} {\bibfnamefont {T.}~\bibnamefont {Lippert}}, \bibinfo {author}
  {\bibfnamefont {B.}~\bibnamefont {Orth}}, \bibinfo {author} {\bibfnamefont
  {P.}~\bibnamefont {Ueberholz}}, \bibinfo {author} {\bibfnamefont
  {K.}~\bibnamefont {Schilling}},\ and\ \bibinfo {author} {\bibfnamefont
  {T.}~\bibnamefont {Struckmann}} (\bibinfo {collaboration} {TXL, T(X)L}),\
  }\bibfield  {title} {\bibinfo {title} {{Static potentials and glueball masses
  from QCD simulations with Wilson sea quarks}},\ }\href
  {https://doi.org/10.1103/PhysRevD.62.054503} {\bibfield  {journal} {\bibinfo
  {journal} {Phys. Rev. D}\ }\textbf {\bibinfo {volume} {62}},\ \bibinfo
  {pages} {054503} (\bibinfo {year} {2000})},\ \Eprint
  {https://arxiv.org/abs/hep-lat/0003012} {arXiv:hep-lat/0003012} \BibitemShut
  {NoStop}%
\bibitem [{\citenamefont {Orth}\ \emph {et~al.}(2005)\citenamefont {Orth},
  \citenamefont {Lippert},\ and\ \citenamefont {Schilling}}]{Orth:2005kq}%
  \BibitemOpen
  \bibfield  {author} {\bibinfo {author} {\bibfnamefont {B.}~\bibnamefont
  {Orth}}, \bibinfo {author} {\bibfnamefont {T.}~\bibnamefont {Lippert}},\ and\
  \bibinfo {author} {\bibfnamefont {K.}~\bibnamefont {Schilling}},\ }\bibfield
  {title} {\bibinfo {title} {{Finite-size effects in lattice QCD with dynamical
  Wilson fermions}},\ }\href {https://doi.org/10.1103/PhysRevD.72.014503}
  {\bibfield  {journal} {\bibinfo  {journal} {Phys. Rev. D}\ }\textbf {\bibinfo
  {volume} {72}},\ \bibinfo {pages} {014503} (\bibinfo {year} {2005})},\
  \Eprint {https://arxiv.org/abs/hep-lat/0503016} {arXiv:hep-lat/0503016}
  \BibitemShut {NoStop}%
\bibitem [{\citenamefont {Wolff}(2004)}]{Wolff:2003sm}%
  \BibitemOpen
  \bibfield  {author} {\bibinfo {author} {\bibfnamefont {U.}~\bibnamefont
  {Wolff}} (\bibinfo {collaboration} {ALPHA}),\ }\bibfield  {title} {\bibinfo
  {title} {{Monte Carlo errors with less errors}},\ }\href
  {https://doi.org/10.1016/S0010-4655(03)00467-3} {\bibfield  {journal}
  {\bibinfo  {journal} {Comput. Phys. Commun.}\ }\textbf {\bibinfo {volume}
  {156}},\ \bibinfo {pages} {143} (\bibinfo {year} {2004})},\ \bibinfo {note}
  {[Erratum: Comput.Phys.Commun. 176, 383 (2007)]},\ \Eprint
  {https://arxiv.org/abs/hep-lat/0306017} {arXiv:hep-lat/0306017} \BibitemShut
  {NoStop}%
\end{thebibliography}%

\end{document}